\UseRawInputEncoding
\documentclass[aps, pra, twocolumn,showpacs,showkeys,superscriptaddress, nofootinbib]{revtex4-2}
\usepackage{graphicx,amsmath,hyperref,url}
\usepackage{xcolor}
\usepackage{braket}
\usepackage{amsmath, amssymb}
\usepackage{caption}
\usepackage{subcaption}
\usepackage{natbib}
\usepackage{soul}
\usepackage{booktabs} 
\usepackage{siunitx}
\usepackage{tabularx} 
\usepackage{makecell} 
\usepackage{multirow}
\usepackage{cellspace}
\begin{document}

\title{High-fidelity two-qubit gates with transmon qubits using bipolar flux pulses and tunable couplers}

\author{Nikita S. Smirnov}
\thanks{Corresponding author, smirnovns@bmstu.ru}
\affiliation{Shukhov Labs, Quantum Park, Bauman Moscow State Technical University, Moscow, 105005, Russia}
\affiliation{Dukhov Automatics Research Institute (VNIIA), Moscow 127030, Russia}
\author{Aleksei R. Matanin}

\author{Anton I. Ivanov}

\author{Vladimir V. Echeistov}

\author{ Nikita D. Korshakov}

\author{Elizaveta I. Malevannaya}
\author{ Viktor I. Polozov}

\author{Bogdan K. Getmanov}
\author{Anastasia A. Solovieva}
\author{Daria A. Moskaleva}
\author{Elizaveta A. Krivko}
\author{Dmitry O. Moskalev}
\author{Dmitry A. Mikhalin}
\author{Igor S. Korobenko}
\author{Denis E. Shirokov}
\affiliation{Shukhov Labs, Quantum Park, Bauman Moscow State Technical University, Moscow, 105005, Russia}

\author{Ilya A. Ryzhikov}
\affiliation{Institute of Theoretical and Applied Electrodynamics, Moscow, 125412, Russia}
\affiliation{Shukhov Labs, Quantum Park, Bauman Moscow State Technical University, Moscow, 105005, Russia}

\author{Alexander V. Andriyash}
\affiliation{Dukhov Automatics Research Institute (VNIIA), Moscow 127030, Russia}

\author{Ilya A. Rodionov}
\thanks{Corresponding author, irodionov@bmstu.ru}
\affiliation{Shukhov Labs, Quantum Park, Bauman Moscow State Technical University, Moscow, 105005, Russia}
\affiliation{Dukhov Automatics Research Institute (VNIIA), Moscow 127030, Russia}

\date{\today}

\begin{abstract}
High-fidelity two-qubit gates are essential for scalable quantum computing. We present a scheme based on superconducting transmon qubits and a control pulse delivery protocol that enables arbitrary controlled-phase gates modulated solely by an independent arbitrary waveform generator pulse. We combined a tunable coupler design with bipolar flux-pulsing to demonstrate a controlled-phase gate, achieving a peak fidelity of 99.5\%. Our gates inherit the advantages of both approaches: minimal residual ZZ coupling, built-in echo-like low-frequency noise protection, and time-scale control pulse reproducibility, while remaining easy to calibrate. We optimize the system’s energy levels to mitigate leakage to the coupler and suppress residual interactions. Numerical simulations of the scheme as three qutrits indicate that an error below $1 \times 10^{-3}$ is achievable. We confirm the scalability potential of the proposed scheme on high-fidelity 4-qubit and 8-qubit quantum processors.

\end{abstract}



\maketitle


\section{INTRODUCTION}
High-fidelity two-qubit  gates are an essential building block for scalable noisy quantum processors \cite{kim2023evidence, Preskill2018} and quantum error correction \cite{Acharya_Abanin_Aghababaie-Beni_Aleiner_Andersen_Ansmann_Arute_Arya_Asfaw_Astrakhantsev__2025}. In superconducting qubit implementation, even though high fidelities have been achieved in developing qubit types such as fluxoniums \cite{moskalenko2022high, simakov2023coupler}, transmon qubits \cite{Koch_Yu_Gambetta_Houck_Schuster_Majer_Blais_Devoret_Girvin_Schoelkopf_2007} have become preferable for multiqubit systems due to their high coherence, controllability, lower crosstalk and connectivity. These same advantages have enabled the creation of multi-qubit arrays for simulating metamaterials \cite{besedin2021topological,cai2019observation} and probing entanglement in quantum many-body systems \cite{karamlou2024probing, neill2016ergodic}.  

One of the widely used high-fidelity gate types \cite{Strauch_Johnson_Dragt_Lobb_Anderson_Wellstood_2003,DiCarlo_Chow_Gambetta_Bishop_Johnson_Schuster_Majer_Blais_Frunzio_Girvin_et_2009} for transmon qubits employs a baseband magnetic flux pulse to bring the computational state $\ket{11}$ into resonance with the non-computational state $\ket{02}$$(\ket{20})$, implementing a CPhase gate. Baseband pulse control enables fast gates with low leakage for weakly anharmonic transmons \cite{Foxen_Neill_Dunsworth_Roushan_Chiaro_Megrant_Kelly_Chen_Satzinger_Barends_2020a,Martinis_Geller_2014a, Barends_Quintana_Petukhov_Chen_Kafri_Kechedzhi_Collins_Naaman_Boixo_Arute_et_2019}. The primary fidelity limitation of this type of gate comes from decoherence induced by flux noise \cite{Schreier_Houck_Koch_Schuster_Johnson_Chow_Gambetta_Majer_Frunzio_Devoret_et_2008,Bylander_Gustavsson_Yan_Yoshihara_Harrabi_Fitch_Cory_Nakamura_Tsai_Oliver_2011, Koch_DiVincenzo_Clarke_2007, Yan_Bylander_Gustavsson_Yoshihara_Harrabi_Cory_Orlando_Nakamura_Tsai_Oliver_2012}. During the gate, the flux control pulse significantly detunes the qubit from its sweet spot (flux-insensitive point) to achieve the resonance $\ket{11}-\ket{02}(\ket{20})$, which increases the decoherence rate. To mitigate the noise, flux control lines are equipped with high-pass bias-tees and low-pass filters. However, such filtering introduces substantial pulse distortions in the control lines \cite{Foxen_Mutus_Lucero_Jeffrey_Sank_Barends_Arya_Burkett_Chen_Chen_et_2019}. The most challenging are long-timescale distortions, as their compensation requires pre-calculated pulse predistortions that are incompatible with real-time operations or extended repeatable waveform lengths - essential requirements for programmable quantum computers.
To overcome this issue there is a known solution enabled by zero-average bipolar flux-pulsing (net-zero (NZ) \cite{Rol_Battistel_Malinowski_Bultink_Tarasinski_Vollmer_Haider_Muthusubramanian_Bruno_Terhal_2019, Andersen_Remm_Lazar_Krinner_Lacroix_Norris_Gabureac_Eichler_Wallraff_2020}), and its later modification set-net-zero (SNZ) \cite{Negîrneac_Ali_Muthusubramanian_Battistel_Sagastizabal_Moreira_Marques_Vlothuizen_Beekman_Zachariadis_et_2021, h7cv-xgw2}. The zero-average gate characteristic prevents long-timescale distortions in flux control lines and simultaneously achieves high-fidelity two-qubit gates. However, previously realized net-zero pulse schemes utilized direct capacitive coupling between the qubits and suffered from high parasitic ZZ-couplings and frequency crowding \cite{Rol_Battistel_Malinowski_Bultink_Tarasinski_Vollmer_Haider_Muthusubramanian_Bruno_Terhal_2019, Negîrneac_Ali_Muthusubramanian_Battistel_Sagastizabal_Moreira_Marques_Vlothuizen_Beekman_Zachariadis_et_2021}, which led to high idling and single-qubit gates errors, either required large frequency detuning between the neighboring qubits \cite{h7cv-xgw2}. The tunable coupling scheme \cite{Yan_Krantz_Sung_Kjaergaard_Campbell_Orlando_Gustavsson_Oliver_2018} is a well-known solution for eliminating residual parasitic interactions that does not require large qubit frequency detuning. We have compiled a detailed overview of flux-activated gates on transmon qubits in Appendix A.

In this work, we combined a tunable coupler design with bipolar flux-pulsing to demonstrate a high-fidelity CPhase gate with cross-entropy benchmarked \cite{Boixo_Isakov_Smelyanskiy_Babbush_Ding_Jiang_Bremner_Martinis_Neven_2018, Arute_Arya_Babbush_Bacon_Bardin_Barends_Biswas_Boixo_Brandao_Buell_2019} performance of 99.48(9)\%. Our hybrid approach inherits advantages from both techniques: minimal residual ZZ coupling ($<5$ kHz) from the tunable coupler, the built-in echo low-f noise protection, and crucially time-scale control pulse reproducibility from the bipolar flux-pulsing. We introduce a gate scheme called the Bipolar Interleaved Amplitude-Tuned (BAT) pulse sequence. The proposed gate scheme features simple calibration and enables arbitrary acquired CPhase angles without additional pulse tuning. Through numerical simulations, we demonstrate two-qubit gates fidelity over $99.9\%$ can be achieved in our scheme with improved pulse short-timescale distortions or coherence. Experimental validation on a four-qubit superconducting quantum processor yielded simultaneous single-qubit gates fidelities exceeding $99.925(6)\%$ and CZ fidelities exceeding  $99.13(0.11)\%$, confirming the proposed scheme's scalability.
\section{RESULTS}
First, we present the setup of a two-qubit device. Next, we demonstrate the experimental realization of the net-zero gate with a tunable coupling strength on the two-qubit device. We then present numerical simulations to evaluate key sources of error. Finally, we show the implementation of the proposed gate scheme on a four-qubit device to validate the scalability of our approach.
\subsection{Device setup}
\begin{figure}
    \centering
    \includegraphics[width=0.95\linewidth]{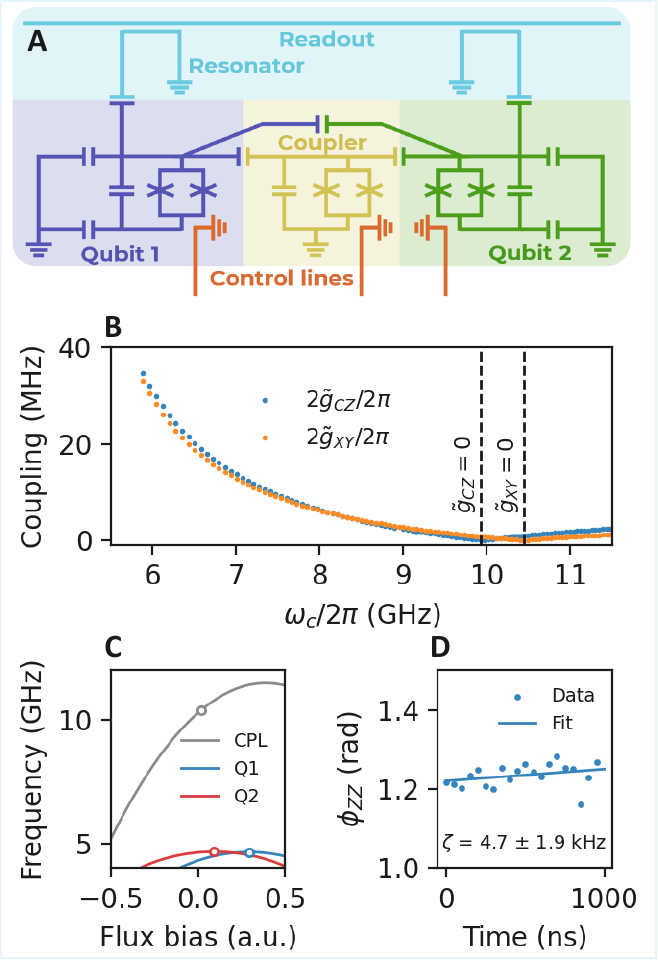}
    \caption{
    \textbf{(a)} Simplified 2-qubit circuit schematic.
    \textbf{(b)} Swap rates $|\widetilde{g}_{CZ}|/2\pi$ and $|\widetilde{g}_{XY}|/2\pi$ extracted by fitting the energy oscillations $\ket{002}$  and $\ket{101}$ , and $\ket{100}$ and $\ket{001}$, respectively, with sinusoidal curves as a function of the coupler frequency. 
    \textbf{(c)} $\ket{0}\xrightarrow{}\ket{1}$ transition frequencies of the qubits (red and blue) and the coupler (gray) as a function of the offset flux bias. Circles indicate idling points. 
    \textbf{(d)} Accumulated ZZ angle at the idle frequencies as a function of the idling time. The uncertainty represents the 68\% confidence interval.
    }
    
    \label{fig:fig1}
\end{figure}

The experimental device of the two-qubit circuit consists of two floating double-padded transmons, serving as data qubits ($Q1$ and $Q2$) and a grounded transmon as a tunable coupler ($CPL$) as described in Ref. \cite{Yan_Krantz_Sung_Kjaergaard_Campbell_Orlando_Gustavsson_Oliver_2018}. The schematic of the device is shown in Fig. \ref{fig:fig1}\textbf{a}. 
The two qubits are capacitively coupled to a coupler with coupling strengths $g_{1c},g_{2c}$ and to each other with a coupling strength $g_{12}$. The qubit-coupler interaction is much stronger than direct qubit-qubit coupling: $g_{1c}\approx g_{2c} \gg g_{12} $. Both qubits and the coupler contain SQUID loops connected to flux control lines for biasing and fast frequency tuning. The two-qubit gates are implemented by dynamically tuning the resonant frequencies of qubit 1 $\omega_1/2\pi$ and the coupler $\omega_c/2\pi$. Qubit 2 is parked at a constant frequency $\omega_2/2\pi$ during the gate. The coupler is detuned from the qubits $\Delta_{jc} \equiv \omega_{j}-\omega_{c}<0$ $(j=1,2)$, while maintaining the dispersive coupling regime  $g_{jc}\ll\Delta_{jc}$ $(j=1,2)$. The qubits are connected to individual microwave control lines for single-qubit X- and Y-rotations, as well as flux-lines for Z-control. The qubits are also dispersively coupled to the coplanar waveguide resonators for quantum state readout. Table \ref{tab:parameters} summarizes the main parameters of our circuit, where the reduced $T_{2R }$ of qubit 1 at the gate operating point is caused by detuning from the sweet spot, leading to increased sensitivity to flux noise, while qubit 2 remains near the sweet spot during the gate and retains its coherence. The qubit's sweet spot frequency could be shifted via post-fabrication critical current adjustment to reduce gate detuning \cite{smirnov2025subangstrom}.  The qubit-coupler-qubit system, approximated as Duffing oscillators, can be described by the Hamiltonian \cite{Yan_Krantz_Sung_Kjaergaard_Campbell_Orlando_Gustavsson_Oliver_2018}:
\begin{flalign}
    \label{eq::H}
    H/\hbar=\sum_i \left(\omega_i b_i^\dagger b_i + \frac{\eta_i}{2} b_i^\dagger b_i^\dagger b_i b_i \right) + \sum_{i<j} g_{ij} (b_i - b_i^\dagger)(b_j - b_j^\dagger),
\end{flalign}
where $b_i^\dagger$ and $b_i$  $(i, j \in \{1, 2, c\})$ are the raising and lowering operators, respectively, defined in the basis of the corresponding oscillator, and $\eta_i$ represents the anharmonicity of the oscillators. 
After the transformation of the Hamiltonian with the coupler decoupling from the system, we obtain the following effective two-qubit Hamiltonian \cite{Zhao_Lan_Xu_Xue_Blank_Tan_Yu_Yu_2021}:
\begin{equation}
    \tilde{H} = \tilde{\omega}_1 ZI/2 + \tilde{\omega}_2 IZ/2 + J (XX + YY)/2 + \zeta ZZ/4,
    \label{eq::Htrans}
\end{equation}
where $(X,Y,Z,I)$ are Pauli and unity operators, the order of the operator in the equation corresponds to the qubit index. $XY$ coupling between the qubits is approximated as $J = g_{12} + g_{1c}g_{2c}/\Delta$, where $1/\Delta = (1/\Delta_{1c} + 1/\Delta_{2c})$.
The tunable coupler is used to either turn off the interaction between the qubits $\widetilde{g}=0$ by setting it to the idle frequency $\omega_c^{off}$, or to activate the interaction $|\widetilde{g}|>0$ by tuning the coupler to the desired frequency. 

This scheme with a tunable coupler exhibits two main error types arising from parasitic interactions: static errors during qubit idling, readout, or execution of single-qubit gates, and dynamic errors from leakage to non-computational states during two-qubit gates. Both error types are solved by careful energy level engineering. We use the notation $\ket{Q1, CPL,Q2}$ to represent the eigenstates of the system.

\begin{table}[htbp]
\centering
\caption{Summary of parameters for two-qubit device}
\label{tab:parameters}
\begin{tabularx}{0.48\textwidth}{lccc}
\toprule
Parameter & Q1 & Q2 & CPL \\
\midrule
$\omega_i/2\pi^a$ (\si{GHz}) & 4.650 & 4.662 & 10.234 \\
$\eta_i/2\pi^b$ (\si{GHz}) & -0.211 & -0.212 & -0.256 \\
$g_{ic}/2\pi$ (\si{MHz}) & 262 & 262 & -- \\
$g_{12}/2\pi$ (\si{MHz}) & 14.91 & -- & -- \\
$\omega_r/2\pi^c$ (\si{GHz}) & 6.328 & 6.277 & -- \\
$\chi/2\pi^d$ (\si{kHz}) & 105 & 90 & -- \\
$T_{1,idle}^e$ (\si{\micro s}) & 55.5 & 111.3 & -- \\
$T_{1,gate}^f$ (\si{\micro s}) & 110.7 & 111.3 & -- \\
$T_{2R,idle}^e$ (\si{\micro s}) & 29.7 & 31.0 & -- \\
$T_{2R,gate}^f$ (\si{\micro s}) & 4.5 & 31.0 & -- \\
$T_{2E,idle}^e$ (\si{\micro s}) & 100.9 & 110.8 & -- \\
$T_{2E,gate}^f$ (\si{\micro s}) & 48.5 & 110.8 & -- \\
\bottomrule
\end{tabularx}
\begin{flushleft}
\begin{tabular}{l}

\small{$^a$ Qubit idle frequency}\\
\small{$^b$ Anharmonicity}\\
\small{$^c$ Readout resonator frequency}\\
\small{$^d$ Readout resonator dispersive shift}\\
\small{$^e$ Both qubits and the coupler parked at their idle frequencies} \\
\small{$^f$ Qubits at their gate frequencies and the coupler at $\omega_c$} \\

\end{tabular}
\end{flushleft}
\end{table}

The first error source type comprises static parasitic ZZ-interaction, which mainly occurs from qubit coupling with high-energy states~\cite{Zhao_Lan_Xu_Xue_Blank_Tan_Yu_Yu_2021,Zhao_Linghu_Li_Xu_Wang_Xue_Jin_Yu_2022, Sung_Ding_Braumüller_Vepsäläinen_Kannan_Kjaergaard_Greene_Samach_McNally_Kim_et_2021, Xu_Chu_Yuan_Qiu_Zhou_Zhang_Tan_Yu_Liu_Li_et_2020}: $\ket{101} \leftrightarrow \ket{020}$ and $\ket{101} \leftrightarrow \ket{200}$ ($\ket{002}$). It is necessary to ensure that the ZZ-interaction and XY-interaction are simultaneously switched off at a single coupler frequency to fully isolate qubits both in resonant and off-resonant coupling regime.

The second type of errors is caused by the implementation of a coupler into the qubit system, since all energy levels associated with the coupler are non-computational and act as leakage channels~\cite{Sung_Ding_Braumüller_Vepsäläinen_Kannan_Kjaergaard_Greene_Samach_McNally_Kim_et_2021, Foxen_Neill_Dunsworth_Roushan_Chiaro_Megrant_Kelly_Chen_Satzinger_Barends_2020a}. To perform fast gates, the coupler is detuned close to qubit frequencies where it is strongly hybridized with qubit levels. The resulting non-adiabatic evolution of the qubit-coupler system, described by Landau-Zener effect, induces leakage errors.

To mitigate such leakages, complex frequency trajectories of both qubits and coupler during gate execution are typically used to ensure optimal adiabaticity~\cite{Sung_Ding_Braumüller_Vepsäläinen_Kannan_Kjaergaard_Greene_Samach_McNally_Kim_et_2021, Martinis_Geller_2014a}. However, this approach complicates calibration procedures, which is undesirable for multiqubit systems. In our design, we instead minimize the overlap between computational and leakage states through engineered energy levels. The overlap between the computational eigenstate vector and the leakage states both for 1-photon subspace ($|\braket{100|010}|^2$, $|\braket{001|010}|^2$) and 2-photon ($|\braket{101|110}|^2$ and $|\braket{101|011}|^2$) is proportional to $(g_{1c(2c)}^2)/(\Delta_{1c(2c)}^2)$. The effective coupling strength between the qubits $\tilde{g}$, assuming that $g_{1c} \approx g_{2c} \gg g_{12}$ and the qubits are weakly detuned, is proportional to $(g_{1c(2c)}^2)/\Delta_{1c(2c)}$. 

We exploited the fact that as the qubit-coupler detuning ($\Delta_{1c(2c)}$) decreases, the effective coupling strength between the qubits $\tilde{g}$ grows faster than the state overlap. We designed our qubits to be strongly coupled to the tunable coupler, while maintaining the coupler at large detuning -- both at its idle frequency $\omega_c^{\text{off}}$, and during two-qubit gates. With our circuit parameters this results in state overlap less than 1\% with the coupling strength between qubits corresponding to 20\,ns $\ket{101} \leftrightarrow \ket{200}$ ($\ket{002}$) swap duration. The chosen system parameters also allow us to find ZZ-interaction free system frequencies. The detailed results of the system parameters optimization are presented in Appendix B.

We experimentally determined the residual ZZ-interaction $\zeta$ at the idling frequencies of the qubits and tunable coupler by measuring the conditional phase of $Q1$ as a function of time when $Q2$ is either excited or not, and found $|\zeta/2\pi| < 5$\,kHz at $\omega_c^{\text{off}}/2\pi = 10.234$\,GHz (Fig. \ref{fig:fig1}\textbf{d}).

\subsection{CPhase implementation}
To implement CPhase gates, we utilize the non-adiabatic transition between $\ket{101}$ and $\ket{002}$. Adjusting the coupler frequency effectively tunes the coupling strength between $\ket{101}$ and $\ket{002}$ (denoted as $\tilde{g}_{\text{CZ}}$). First, we experimentally measured the effective coupling strength $\tilde{g}_{\text{CZ}}$ by observing the energy exchange between $\ket{101}$ and $\ket{002}$ as a function of the coupler frequency. We prepared the $\ket{101}$ state, rapidly tuned the frequency of $Q1$ to bring $\ket{101}$ and $\ket{002}$ into resonance, and activated the coupling $\tilde{g}_{\text{CZ}}$ by shifting the coupler frequency $\omega_c$. We waited for a variable time and measured the population of $\ket{101}$. We repeated the experiment for various coupler frequencies $\omega_c$. In Fig.\ref{fig:fig1}\textbf{b} we plot the effective coupling strength $\tilde{g}_{\text{CZ}}$ as a function of the coupler frequency by fitting the energy exchange oscillations and demonstrate a maximum coupling strength exceeding $\tilde{g}_{\text{CZ}} > 40$ MHz at the coupler detuning $> 1$ GHz.

\begin{figure*}
    \centering
    \includegraphics[width=0.95\linewidth]{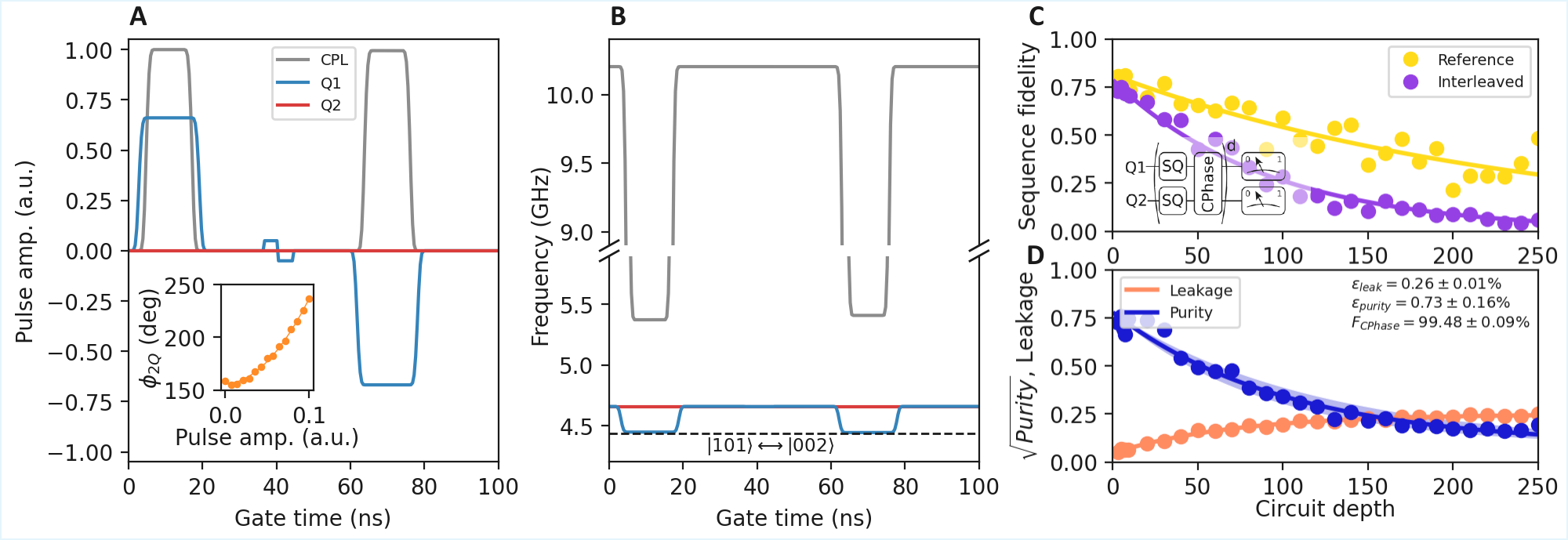}
    \caption{
    \textbf{(a)} Experimentally implemented BAT pulses. Control pulses, plotted over time (horizontal axis), expressed as the flux applied to each transmon qubit and the coupler. The inset shows acquired CPhase angle as a function of the weak bipolar flux pulse amplitude $a_\text{int}$. 
    \textbf{(b)} Qubit and coupler transition frequencies as a function of gate time. Dashed line indicates qubit interaction frequency.
    \textbf{(c)} The results of interleaved XEB for the 100-ns-long CPhase gate, consisting of two 20-ns pulses.  
    \textbf{(d)} Accumulation of population in the leakage subspace $\{\vert 002\rangle, \vert 102\rangle, \vert 202\rangle\}$ and decay of square root of purity during interleaved XEB. We extract average XEB reference error $0.29(2)\%$, interleaved cycle error $0.81(3)\%$, leakage cycle error $0.26(1)\%$ and purity cycle error $0.73(0.16)\%$. The uncertainty represents the 68\% confidence interval. 
    }
    
    \label{fig:fig2}
\end{figure*}

The ideal CPhase gate for a two-qubit system with a tunable coupler is described by the unitary transformation:
\begin{equation}
U = \begin{pmatrix}
1 & 0 & 0 & 0 \\
0 & e^{i\phi_{001}} & 0 & 0 \\
0 & 0 & e^{i\phi_{100}} & 0 \\
0 & 0 & 0 & e^{i\phi_{101}}
\end{pmatrix}
\end{equation}
in the computational basis $\{\ket{000}, \ket{001}, \ket{100}, \ket{101}\}$, where $\phi_{001}$ and $\phi_{100}$ are single-qubit phases, and $\phi_{101}$ is a two-qubit phase.

The CPhase gate employing the proposed BAT scheme with bipolar net-zero pulses and a coupler is realized in three steps (Fig. \ref{fig:fig2}). First, $\ket{101}$ is brought into resonance with $\ket{002}$ using a Gaussian smoothed pulse with amplitude $+V$ and duration $t_p$, while simultaneously activating the $\tilde{g}_{\text{CZ}}$ coupling by detuning the coupler frequency, and then returned, to complete the full population exchange between $\ket{101}$ and $\ket{002}$. Next, with coupling turned off at $\omega_c^{\text{off}}$ after a short delay, a weak bipolar pulse with arbitrary amplitude $\pm a_{\text{int}}$ is applied to one of the qubits to adjust the desired conditional phase defined by $\phi_{2Q} = \phi_{101} - \phi_{001} - \phi_{100}$. 

Finally, another strong pulse with negative amplitude $-V$ is applied to the qubit while turning on the coupling, returning the excitation from $\ket{002}$ back into the computational subspace. 

Notably, only unipolar pulses are applied to the coupler, as it does not require high coherence. The key advantage of bipolar-pulse gates with a coupler over conventional adiabatic gates with directly coupled qubits lies in their straightforward calibration of arbitrary acquired CPhase angle $\phi_{2Q}$ and zero leakage to $\ket{002}$. 

Thus, the calibration protocol for BAT pulses involves only two key parameters: a pulse amplitude $\pm V$ crucial for the complete $\ket{100} \leftrightarrow \ket{002}$ transition, and a weak intermediate pulse amplitude $\pm a_{\text{int}}$ defining the conditional phase $\phi_{2Q}$. The intermediate pulse does not interfere with the previous and subsequent exchange pulses and can be calibrated independently. Unlike SNZ~\cite{Negîrneac_Ali_Muthusubramanian_Battistel_Sagastizabal_Moreira_Marques_Vlothuizen_Beekman_Zachariadis_et_2021, h7cv-xgw2}, our approach does not require gate duration adjustment to achieve different acquired CPhase angles. When considering real arbitrary waveform generator (AWG) limitations, amplitude tuning while keeping the timing fixed is preferable due to limited sampling rate and high pulse amplitude resolution. The proposed method enables full in-situ calibration of a continuous set of two-qubit gates, parameterized by the amplitude of the intermediate pulse.

We experimentally implemented a bipolar-pulse gate with tunable coupling, using Gaussian-smoothed pulses of duration $t_p = \SI{20}{\nano\second}$ and an effective coupling strength $\tilde{g}_{\text{CZ}} = \SI{40}{\mega\hertz}$. To suppress residual pulse distortions, we introduced a delay $t_d = \SI{20}{\nano\second}$, after both strong and weak pulses. By sweeping the intermediate pulse amplitude $|a_{\text{int}}|$ we reconstructed the full set of achievable CPhase angles $\phi_{\text{2Q}}$ using quantum state tomography (the inset in Fig. \ref{fig:fig2}\textbf{a}). The total duration of the small bipolar pulse is 8 ns.

We characterized the BAT CPhase gate performance using cross-entropy benchmarking (XEB)~\cite{Boixo_Isakov_Smelyanskiy_Babbush_Ding_Jiang_Bremner_Martinis_Neven_2018, Arute_Arya_Babbush_Bacon_Bardin_Barends_Biswas_Boixo_Brandao_Buell_2019}. We performed sequences of random single-qubit gates for reference and sequences with interleaved CPhase (each cycle consists of a random single
qubit gate applied
to each qubit followed by the CPhase gate) at variable circuit depths $d$. For each circuit depth, we sample over 30 different random circuits.The probability distributions were obtained from 4096 single-shot readouts. For gate calibration and benchmarking, we used 3-state readout by applying a Gaussian mixture model for classification. Readout correction techniques were not applied. We extracted depolarization fidelity $p$ approximated as $Ap^d + B$, where $A$ and $B$ absorbed state preparation and measurement errors (Fig. \ref{fig:fig2}\textbf{c}). We implemented single-qubit gates using \SI{40}{\nano\second}-long microwave pulses. We extracted CPhase gate fidelity through the formula:
\begin{equation}
    F = p + \frac{1 - p}{D},
\end{equation}
where $p = p_2/p_1$ with respectively reference depolarization fidelity $p_1$ and interleaved sequence fidelity $p_2$, $D$ is Hilbert space dimension. We measured a CPhase gate fidelity of $99.48(9)\%$ for the best-fit unitary with the CPhase angle $\phi_{2Q}=160 \deg$. The uncertainty represents the 68\% confidence interval throughout.

To disentangle error contributions of the interleaved cycle, we additionally quantified leakage error $\epsilon_{\text{leak,int}}$ per-cycle purity error~\cite{Wallman_Granade_Harper_Flammia_2015} $\epsilon_{\text{purity,int}}$. The data in Fig. \ref{fig:fig2}\textbf{d} reveal per-cycle purity error of $0.73(0.16)\%$ and per-cycle leakage error of $0.26(1)\%$. Coherent error given as $\epsilon_{\text{XEB,int}} - \epsilon_{\text{purity,int}}$ is about $0.08 (0.2)\%$ per cycle. The CPhase cycle error is dominated by decoherence quantified as $\epsilon_{\text{purity,int}} - \epsilon_{\text{leak,int}}$ is about $0.47(0.17)\%$. The leakage error for reference gate sequences was negligible and can be effectively considered zero. At the same time, the measured reference average purity error $\epsilon_{\text{purity,ref}}$ is $0.29(1)\%$.
Taking into account all the delays included in the pulse scheme, the decoherence error of the CPhase gate can be quantified as:
\begin{equation}
    \epsilon_{\text{decoh,CPhase}} = \epsilon_{\text{purity,int}} - \epsilon_{\text{leak,int}}  - \epsilon_{\text{purity,ref}}.
\end{equation}
The estimated decoherence error of 100 ns BAT CPhase gate  is $0.18(0.18)\%$

Assuming that the \SI{20}{\nano\second} delay after the CPhase gate is not noise protected by the echo effect, it may have the same decoherence rate as a reference single-qubit cycle. Then, the decoherence error of the ideal BAT CPhase gate without delays can be estimated using the following formula:
\begin{equation}
    \epsilon_{\text{decoh,CPhase}} = \epsilon_{\text{purity,int}} - \epsilon_{\text{leak,int}} - 1.5 \times \epsilon_{\text{purity,ref}},
\end{equation}
where coefficient 1.5 solves to add 20 ns delay purity error to the 40 ns reference cycle. Then the estimated decoherence error of the CPhase gate without delay is $0.03(0.18)\%$.

\subsection{Leakage and decoherence contribution to gate errors}
We numerically simulated coherent leakage in BAT CPhase gates and the influence of decoherence on the fidelity. We considered the dynamics of a three-body system as three qutrits. The parameters of the simulated system were the same as the parameters of the experimental circuit.

\begin{figure}
    \centering
    \includegraphics[width=0.95\linewidth]{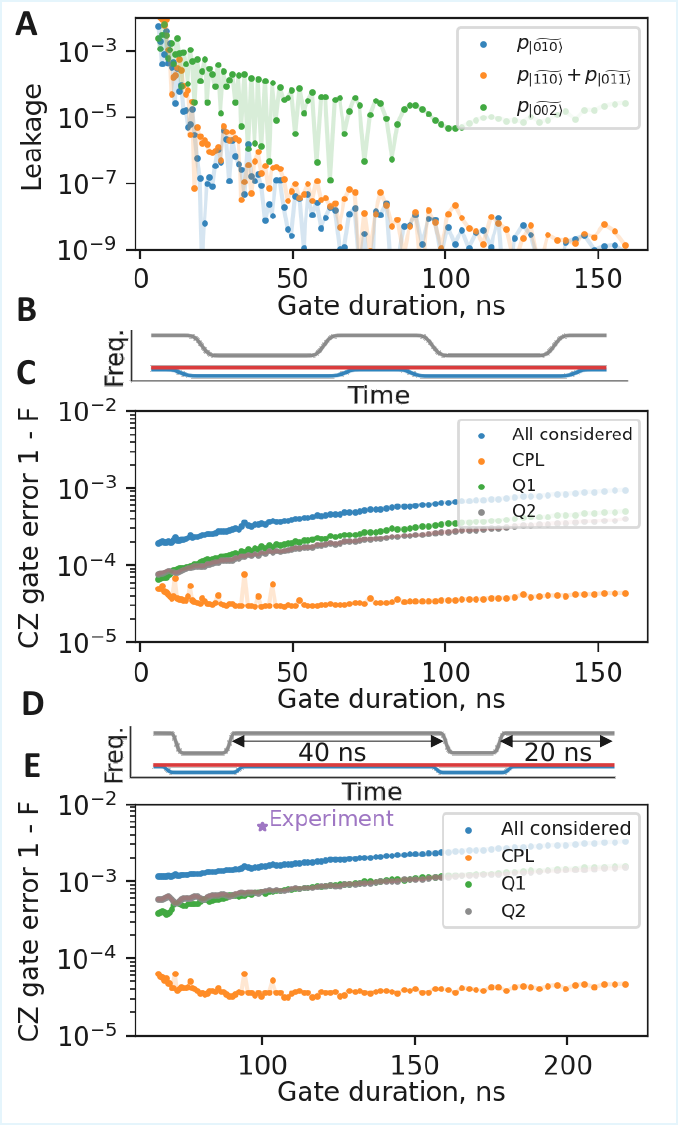}
    \caption{  
    \textbf{(a)} Numerical simulation of coherent leakage of CPhase gates. We prepare $\widetilde{\ket{101}}$ or $\widetilde{\ket{100}}$ for single photon and two-photon manifold respectively, apply control pulse, and then measure the leakage state populations. Leakage to coupler-associated states below $10^{-6}$ is achievable with 20-ns Gaussian-smoothed pulses as used in the experiment. Gate duration is a sum of two opposite-polarity control pulses durations $2t_p$, representing the ideal case without additional delays. 
    \textbf{(b)} Frequency transitions as a function of time used for the simulations of the ideal gate without delays.  \textbf{(c)} Numerical simulations of CPhase errors $1-F$ as a function of gate duration for ideal gate without delays.  
    \textbf{(d)} Frequency transitions as a function of time used for the simulations of the realistic CPhase gate (including delays).
    \textbf{(e)} Numerical simulations of realistic CPhase gate errors.} We prepared 16 states, applied control pulses and calculated the average fidelity from all initial states. Blue, green, gray and orange dots correspond to the cases when noise was in all elements, qubit 1, qubit 2 and tunable coupler, respectively. The purple star represents the experimental value of the total gate error, enabling direct comparison with theoretical performance limits.
    
    \label{fig:fig3}
\end{figure}

First, we found leakage into the coupler-associated leakage states ($\ket{\widetilde{010}}$, $\ket{\widetilde{110}}$, $\ket{\widetilde{011}}$) and into the state $\ket{\widetilde{002}}$ using time-dependent Hamiltonian simulations in QuTiP~\cite{Johansson_Nation_Nori_2013} for the single control pulse (Fig. \ref{fig:fig3}\textbf{a}). In this section, tildes are used because the simulations were performed in the dressed basis. They are omitted in the rest of the text to avoid overloading it. For single-photon excitation, we prepared the initial states $\ket{\widetilde{100}}$ and $\ket{\widetilde{101}}$ for the two-photon case in the system configuration with the coupling turned off at $\omega_c^{\text{off}}$. We simulated a Gaussian-smoothed pulse matching the experimental one, with variable pulse duration. For a given waveform $\omega_1(t)$ and $\omega_c(t)$, we modulated the frequency-dependent parameters $g_{1c}(t)$, $g_{2c}(t)$, and $g_{12}(t)$. After applying the control pulse, we plotted the leakage state populations $p_{\ket{\widetilde{002}}}$, $p_{\ket{\widetilde{010}}}$ and $p_{\ket{\widetilde{110}}} + p_{\ket{\widetilde{011}}}$. Since the leakage states remain weakly hybridized throughout the entire gate duration, the leakage population in the coupler remains below $10^{-5}$ for pulses of 20\,ns and longer. Leakage into the coupler becomes negligible compared to the leakage to $\ket{\widetilde{002}}$. The remaining leakage to $\ket{\widetilde{002}}$ in simulations is due to imperfect swap involving $\ket{\widetilde{101}}$ and requires more precise calibration to reduce it further.

Next, we performed simulations to evaluate the contributions of relaxation and dephasing to gate errors. We executed a full gate set of two pulses with duration $t_p$. The time evolution was calculated by solving the Lindblad master equation:

\begin{equation}
\begin{split}
    \dot{\rho}(t) = -\frac{i}{\hbar} [H(t), \rho(t)] \\ + \sum_{j,i} \left( c_{j,i}(t) \rho(t) c_{j,i}^\dagger - \frac{1}{2} \{ c_{j,i}^\dagger(t) c_{j,i}(t), \rho(t) \} \right),
\end{split}    
\end{equation}
where $\rho$ is the system's density matrix, and $c_{j,i}$ are jump operators defined as in Ref.~\cite{Negîrneac_Ali_Muthusubramanian_Battistel_Sagastizabal_Moreira_Marques_Vlothuizen_Beekman_Zachariadis_et_2021}. For our model, we defined the pure dephasing time of the qubit~1 as
$T_{\phi,1}^E = \left( \frac{1}{T_{2,1}^E} - \frac{1}{2T_{1,1}} \right)^{-1},$
assuming that the built-in echo effect of bipolar pulses suppresses low-frequency noise. To determine the fidelity, we use the principle of quantum process tomography. First, we prepare $N$ states from the set $\{ \ket{0}, \ket{1}, \ket{+}, \ket{-} \} \otimes \{ \ket{0}, \ket{1}, \ket{+}, \ket{-} \}$ and simulate state evolution for both noisy and noiseless cases for each $k$-th state, obtaining $\rho_{\text{noisy}}^k(t_p)$ and $\rho_0^k(t_p)$ respectively for various $t_p$. Then we calculated the fidelity using the formula \cite{Jozsa_1994}:

\begin{equation}
    F = \frac{1}{N} \sum_{k=1}^N tr \left[ \sqrt{ \sqrt{\rho_{\text{noisy}}^k(t_p)} \rho_0^k(t_p) \sqrt{\rho_{\text{noisy}}^k(t_p)}}  \right]^2.
\end{equation}

We first ran simulations using our circuit parameters from Table~1 and obtained a decoherence-induced error of $1 - F = 3.4 \times 10^{-4}$ for 20-ns pulses, corresponding to an ideal 40 ns CPhase gate without delays. This result compares well with the experimental estimate of $3.3 \times 10^{-4}$ for the decoherence-induced error of the ideal gate. We then simulated the gates corresponding to our experimental pulse sequence, which consisted of pulses with an interleaved 40 ns delay and a 20 ns delay after the gate. The simulated decoherence error for the 100 ns gate (including delays) was $1.6 \times 10^{-3}$, in good agreement with the experimentally estimated value of $1.8 \times 10^{-3}$.

Since we were unable to directly measure the coherence of the tunable coupler, we assumed $T_{1,c} = T_{\phi,c}^E = 1\,\mu\text{s}$. We simulated the case where noise was applied only to the coupler and observed an order-of-magnitude lower error compared to the fully noisy model, indicating the coupler's negligible contribution to total decoherence. 

The results obtained can also be compared to the analytical estimate of the decoherence-induced error \cite{chu2021coupler, marxer2023long}:

\begin{equation}
\epsilon_{\text{decoh,CPhase}}= \frac{2}{5} \sum_{i=1,2}  \left[ 
\frac{2t_{p}}{T_{1,i}^{gate}} + 
\frac{2t_{p}}{T_{\phi, i}^{gate}} +  
\frac{t_{d}}{T_{1,i}^{idle}} + 
\frac{t_{d}}{T_{\phi, i}^{idle}}
\right ].
\end{equation}
For our CPhase gate with $t_p = 20$~ns, using considerations as in the text above, we obtained a decoherence error of $2.9 \times 10^{-3}$ with delays $t_d = 60$~ns and $9.9 \times 10^{-4}$ without delays.

The results demonstrate that fidelity is primarily limited by imperfections in control pulses, which require additional delays that increase the decoherence error. 99.9\% fidelity is achievable through optimized pulse shaping and reduced delays.

\subsection{Scalability}
\begin{figure}
    \centering
    \includegraphics[width=0.95\linewidth]{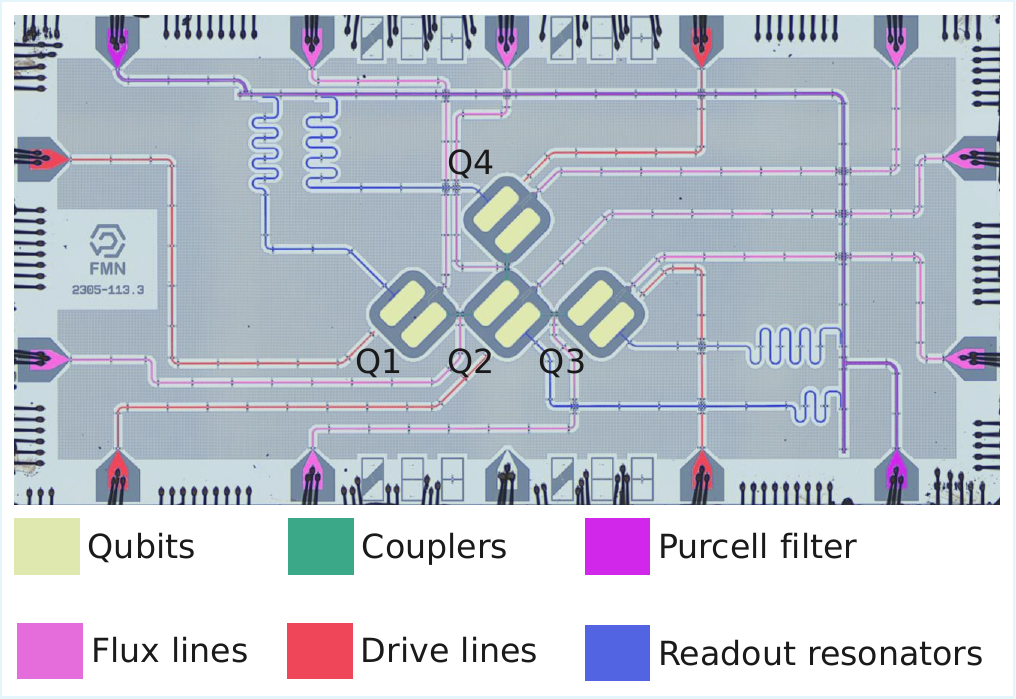}
    \caption{ False-color micrograph of the 4-qubit processor device used to test bipolar control flux pulses approach. Each transmon qubit consists of capacitive pads (yellow) and connected to each other with tunable coupler (green). Each qubit is coupled to a drive line (red) and readout resonators (blue). Fast flux control is performed by individual flux lines (pink). Readout resonators are inductively coupled to the common Purcell filter (violet). 
    }
    
    \label{fig:fig4}
\end{figure}

To demonstrate the scalability of the proposed architecture and the proposed BAT pulsing method for implementing two-qubit gates using bipolar pulses with tunable coupling, we present a 4-qubit quantum processor with minor design adjustments (Fig. \ref{fig:fig4}). Each qubit is connected to an individual microwave control line. The 25-ns single-qubit gates are precisely calibrated using the Derivative Removal by Adiabatic Gate (DRAG) method~\cite{Motzoi_Gambetta_Rebentrost_Wilhelm_2009, Lucero_Kelly_Bialczak_Lenander_Mariantoni_Neeley_O’Connell_Sank_Wang_Weides_et_2010}. Both qubits and couplers are connected to dedicated flux control lines. Readout is performed via quarter-wave resonators coupled to a bandpass Purcell filter, following the design described in Ref.~\cite{matanin2023toward}. Table \ref{tab:processor} summarizes the processor parameters and performance metrics. All benchmarking was performed with all qubits at their idle frequencies. 

\begin{table}[htbp]
\centering
\caption{Performance parameters for the 4-qubit superconducting processor}
\label{tab:processor}
\begin{tabularx}{0.48\textwidth}{lcccc}
\toprule
Parameter & Q1 & Q2 & Q3 & Q4 \\
\midrule
$\omega_i/2\pi^a$ (GHz) & 4.668 & 4.866 & 4.655 & 4.639 \\
$\eta_i/2\pi^b$ (GHz) & -0.235 & -0.232 & -0.245 & -0.241 \\[2pt]
$\zeta/2\pi$ (kHz) & \shortstack{$28$ \\ (Q1-Q2)}   & -- & \shortstack{ $30$ \\ (Q3-Q2)}& \shortstack{ $18$ \\ (Q4-Q2)}\\

\midrule
 $F_{1Q,iso}^c$ (\%) & $99.951(4)$ & $99.936(8)$ & $99.939(5)$ & $99.897(9)$ \\
 $F_{1Q,simult}^d$ (\%) & $99.944(3)$ & $99.925(6)$ & $99.927(7)$ & $99.929(5)$ \\
\midrule
 $F_{CZ}^e$ (\%) & \shortstack{$99.34(8)$ \\ (Q1-Q2)}   & -- & \shortstack{ $99.13(0.11)$ \\ (Q3-Q2)}& \shortstack{ $99.31(9)$ \\ (Q4-Q2)}\\
\midrule
 $F_{R0}^f$ (\%) & 99.08 & 98.85 & 99.17 & 99.62 \\
 $F_{R1}^g$ (\%) & 95.07 & 93.05 & 93.41 & 96.75 \\
\bottomrule
\end{tabularx}
\begin{flushleft}
\begin{tabular}{lcccc}

\small{$^a$ Qubit idle frequency.} & & & & \\
\small{$^b$ Anharmonicity.} & & & & \\
\small{$^c$ Single qubit gates fidelity (isolated).} &     & & & \\
\small{$^d$ Single qubit gates fidelity (simultaneous).} & & & & \\
\small{$^e$ CZ gate fidelity.} & & & & \\
\small{$^f$ Two-state} readout fidelity for preparing $|0\rangle$. & & & & \\
\small{$^g$ Two-state} readout fidelity for preparing $|1\rangle$. & & & & \\

\end{tabular}
\end{flushleft}
\end{table}

We implement CZ gates with a conditional phase $\phi_{\mathrm{2Q}} = 180^\circ$, using pulse duration $t_p = 20$~ns and intermediate delay $t_d = 20$~ns. An additional 20-ns delay is introduced after CZ gates to eliminate residual pulse distortions. All pairs of qubits demonstrate gate fidelities $\mathcal{F} > 99.1\%$, with the highest measured fidelity reaching $99.34(8)\%$. The residual $ZZ$ coupling across the device remains below 30~kHz, enabling simultaneous single-qubit gates fidelities  ranging from $99.925(6)\%$ to $99.944(3)\%$. The gate fidelities on the 4-qubit processor are lower than on the two-qubit device due to higher decoherence-induced error. Across the entire 4-qubit chip, the average $T_{1}$ is 34.5 \si{\micro s} and $T_{1E}$ is 34.0 \si{\micro s}.

\section{CONCLUSION}

In summary, we have developed and implemented a two-qubit unit cell architecture using bipolar pulses and tunable coupling between transmon qubits, enabling high-performance CPhase gates with 99.48(9)\% fidelity. The key advantage of this configuration is threefold: straightforward BAT pulse sequence calibration for arbitrary acquired phases, suppression of residual ZZ interactions at small qubit detunings, while providing immunity against long-timescale flux control pulse distortions. 

Unlike previous implementations of diabatic gates, our scheme requires only two calibration steps, with the CPhase angle being inherently independent of leakage-minimizing pre-calibration. We engineered the energy level structure to simultaneously suppress static errors from residual ZZ coupling below 5\,kHz and minimize leakage to non-computational states associated with the tunable coupler. This design enables fast operations using simple baseband control pulses.

Our results demonstrate that the fidelity is primarily limited by decoherence and can be readily improved through either enhanced coherence times or elimination of short-timescale pulse distortions. Leveraging all advantages of the proposed architecture, we demonstrated its scalability on a 4-qubit processor, achieving average CZ-gate fidelity 99.26\%, peak CZ-gate fidelity 99.34(8)\% and residual ZZ coupling suppression below 30\,kHz. 

\begin{acknowledgments}
Devices were fabricated at the BMSTU Nanofabrication Facility (Functional Micro/Nanosystems, FMNS REC, ID 74300). 

\appendix
\section{Overview of flux-activated gate performance on transmon qubits}
\label{app:A} 

In recent years, the quantum computing community has achieved a substantial number of results in the implementation of high-fidelity quantum gates, using various types of control pulses, qubit types, and their coupling architectures. Here, we focus on transmon qubits, as they have shown the greatest progress in scaling up systems. We have aimed to compile a quantitative comparison of key parameters including the fidelity of two-qubit and single-qubit gates, fidelity coherence limit and residual ZZ-interaction for flux-activated gate implementations and summarized in Table \ref{tab:overview}.  

 We separately highlight the filtering of qubit control lines to emphasize the effectiveness of using bipolar pulses in certain schemes, as well as frequency detuning between qubits $\Delta$ as a method to reduce residual ZZ-interaction. We also include the qubit-coupler coupling strength $g_{qc}$, as a parameter that determines the minimum gate duration and limits the fidelity of single-qubit operations when performed simultaneously  \cite{marxer2025above}. 

For this table, we selected the best reported results from the literature for two-qubit gate fidelity and durations $t_{g}$, average single-qubit gate fidelities (simultaneously performed), used flux-control pulse, type of qubit architecture (frequency-tunable or fixed) and coupler as well as the reported decoherence-limited bound on the two-qubit gate fidelity $F_{2Q,dec}$. When fixed-frequency transmons are used, the pulse shape refers to the coupler. In all other cases, the same pulse shapes are applied to both the qubit and the coupler.

\begin{table*}

    \centering
    
    \caption{An overview of the performance and architecture of two-qubit cells on transmon qubits with flux-activated two-qubit gates.}
    \label{tab:overview}
    \begin{tabular*}
    {\textwidth}{@{\extracolsep{\fill}}lccccccccccc@{}}
\toprule
        Ref.& Gate & Pulse & $t_{g}$ (ns) & \makecell{$F_{2Q}$ \\ $(\%)$ } & \makecell{$F_{1Q}$ \\$(\%)$} & \makecell{ $\zeta$ \\(kHz) }&  \makecell{Coupling\\type} &  \makecell{ $g_{qc}$\\ (MHz) }& \makecell{$F_{2Q,dec}$ \\ $(\%)$ }& \makecell{Flux \\filtering} & \makecell{$\Delta$ \\(MHz)} \\
\midrule 
        \multicolumn{5}{l}{Fixed-frequency qubits with tunable coupling}\\[5pt]

        TUM \cite{glaser2025closed} &CZ& \makecell{Adiabatic \\unipolar \\(Fourier)} &64& 99.91 & 99.815 & 20  & \makecell{Tunable\\ transmon\\coupler} & 64 & N\textbackslash A & N\textbackslash A & 464 \\[15pt]
        
        Riken QC \cite{li2024realization} &CZ& \makecell{Adiabatic \\unipolar \\(Slepian)} &48& 99.90 & 99.984 & 6.3  & \makecell{Double\\ transmon\\coupler} & 100-200 & 99.92 & N\textbackslash A & 400\\[15pt]
        
        IBM \cite{stehlik2021tunable} &CZ& \makecell{Adiabatic \\unipolar \\(custom)} &46& 99.85 & 99.957 & 5  & \makecell{Tunable\\ transmon\\coupler} & 110 & 99.873 & N\textbackslash A & 300 \\[15pt]

        \multicolumn{5}{l}{Flux-tunable qubits with tunable coupling}\\[5pt]
        
        IQM \cite{marxer2025above} & CZ& \makecell{Adiabatic \\unipolar \\(Slepian)} &33 & 99.938 & 99.984 & N\textbackslash A  & \makecell{Long tunable\\ transmon\\coupler} & 66 & 99.935 & \makecell{LPF (1 GHz),\\ Eccosorb} & $<300$ \\[15pt] 

        IQM \cite{marxer2023long} &CZ& \makecell{Adiabatic \\unipolar \\(Slepian)} &33& 99.81 & 99.88 & 2  & \makecell{Long tunable\\ transmon \\coupler} & 53 & 99.83 & LPF (20 kHz) & 200\\[15pt]

        MIT \cite{Sung_Ding_Braumüller_Vepsäläinen_Kannan_Kjaergaard_Greene_Samach_McNally_Kim_et_2021} &CZ& \makecell{Adiabatic \\unipolar \\(Slepian)} &60& 99.76 & 99.92 & 1  & \makecell{Tunable\\ transmon\\coupler} & 72 & 99.83 & LPF (300 MHz) & 160\\[15pt]
        
        Google \cite{Foxen_Neill_Dunsworth_Roushan_Chiaro_Megrant_Kelly_Chen_Satzinger_Barends_2020a} &CZ& \makecell{Adiabatic \\unipolar\\(rectangular)} &13& 99.59 & 99.925 & N\textbackslash A  & \makecell{Tunable\\transmon\\coupler} & N\textbackslash A & 99.96 & LPF (500 MHz) & $<500$\\[15pt]
        
        SKLQSE \cite{Xu_Chu_Yuan_Qiu_Zhou_Zhang_Tan_Yu_Liu_Li_et_2020} &CZ& \makecell{Adiabatic \\unipolar \\(cosine)} &30& 99.48 & 99.87 & 20  & \makecell{Tunable\\ transmon\\coupler} & 112 & 99.6 & LPF & $>600$\\[15pt]
        
        This work &CPhase& \makecell{Adiabatic \\net-zero\\(Gauss)} &100& 99.48 & 99.85 & 5  & \makecell{Tunable\\transmon\\coupler} & 262 & 99.9 & \makecell{Bias-T,\\ LPF (1 GHz),\\ Eccosorb} & 10\\

        Rigetti \cite{field2024modular} &CZ& Parametric &56& 99.13 & 99.7 & 17  & \makecell{Tunable\\transmon\\coupler} & 94 & 99.1 & N\textbackslash A &  N\textbackslash A\\[15pt]

        ETH Zurich \cite{collodo2020implementation} &CZ& \makecell{Adiabatic \\unipolar \\(Gauss flat-top)} &38& 97.9 & 99.73 & 60  &  \makecell{Tunable\\ transmon\\coupler} & 270 & $>99.0$ & \makecell{Bias-T, \\LPF (10 kHz),\\ Eccosorb} & 360 \\[15pt]

        \multicolumn{5}{l}{Flux-tunable qubits with fixed coupling}\\[5pt]

        QuTech \cite{Negîrneac_Ali_Muthusubramanian_Battistel_Sagastizabal_Moreira_Marques_Vlothuizen_Beekman_Zachariadis_et_2021} &CZ& \makecell{Adiabatic \\set net-zero}  &43& 99.54 & \makecell{96 \\(idle)}&893  & \makecell{Fixed\\ resonator} & N\textbackslash A & 99  & N\textbackslash A & $>500$\\[15pt]

        Google \cite{Barends_Quintana_Petukhov_Chen_Kafri_Kechedzhi_Collins_Naaman_Boixo_Arute_et_2019} &CPhase& \makecell{Adiabatic \\unipolar\\(trapezoidal)} &28& 99.42 & 99.85 & N\textbackslash A & \makecell{Direct \\ capacitive} & N\textbackslash A & 99.61 & N\textbackslash A & $>1000$\\[15pt]

        ETH Zurich \cite{h7cv-xgw2} &CZ& \makecell{Adiabatic \\net-zero \\(Gauss)} &78& 99.39 & 99.94 & 4.4  & \makecell{Fixed\\ resonator} & 460 & N\textbackslash A & \makecell{Bias-T, \\LPF (780 MHz),\\ Eccosorb } & $>2000$\\[15pt] 
        
        QuTech \cite{Rol_Battistel_Malinowski_Bultink_Tarasinski_Vollmer_Haider_Muthusubramanian_Bruno_Terhal_2019} &CZ& \makecell{Adiabatic \\net-zero \\(square)} &40& 99.1 & N\textbackslash A & N\textbackslash A  & \makecell{Fixed\\ resonator} & N\textbackslash A & 99.18 & N\textbackslash A & 870\\[15pt]

\bottomrule
    \end{tabular*}
    
    \label{tab:placeholder}
\end{table*}

\section{Optimization of system parameters}

\label{app:B}

\begin{figure*}
    \centering
    \includegraphics[width=1\linewidth]{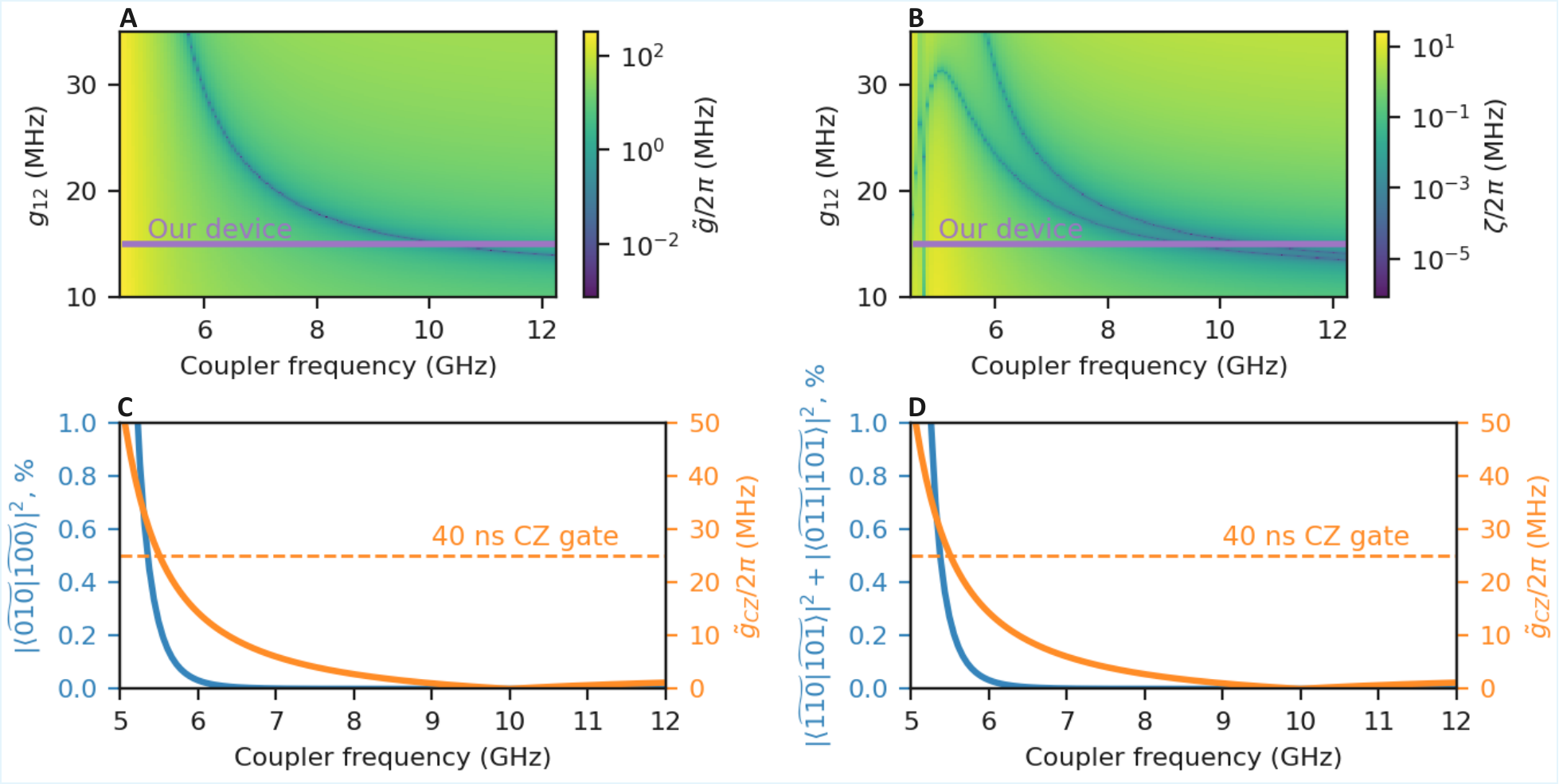}
    \caption{Numerical simulations of
    \textbf{(a)} coupling strength $\tilde{g}$ and
    \textbf{(b)} residual ZZ coupling strength $\zeta$ versus the coupling strength $g_{12}$ and the coupler frequency.
    \textbf{(c)} The state overlap $|\langle100|010\rangle|^2$ in the single-excitation manifold (blue curve) and the coupling strength $\tilde{g}_{\text{CZ}}$ (orange curve) as a function of the coupler frequency.
    \textbf{(d)} The state overlap ($|\langle101|110\rangle|^2 + |\langle101|011\rangle|^2$) in the double-excitation manifold (blue curve) and the coupling strength $\tilde{g}_{\text{CZ}}$ (orange curve) as a function of the coupler frequency.
    }
    
    \label{fig:fig5}
\end{figure*}

\begin{figure}
    \centering
    \includegraphics[width=0.85\linewidth]{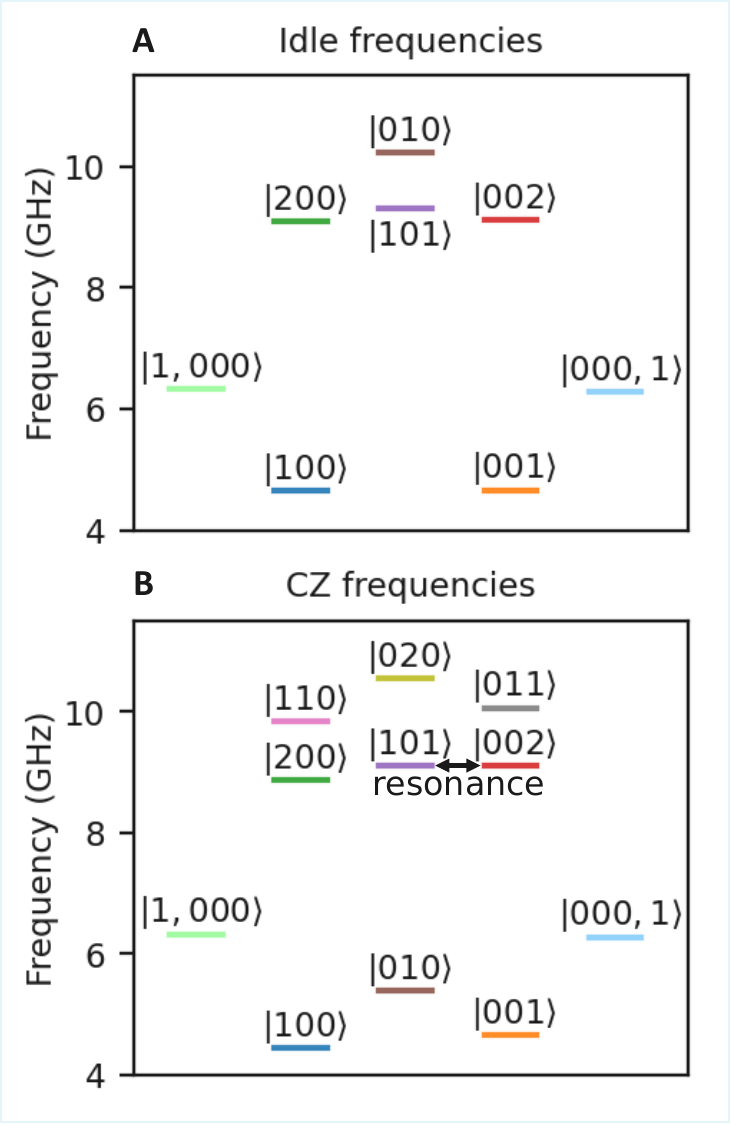}
    \caption{ Energy level diagrams of the 2-qubit device: 
    \textbf{(a)} when the qubit and coupler are parked at their idle frequencies, and 
    \textbf{(b)} during the execution of a CZ gate. $n$ in $\ket{n,Q1,CPL,Q2}$ and $\ket{Q1,CPL,Q2,n}$ denotes energy levels of levels of first and second qubit's resonator respectively. In both configurations, the energy levels associated with leakage through the coupler remain highly detuned (by more than 1 GHz) from the computational states. Furthermore, when parked at idle frequencies, the $\ket{020}$ is highly detuned, and the $\ket{100}-\ket{010} (\ket{001}-\ket{010})$ detunings are high. This combination simultaneously suppresses both ZZ and XY interactions. 
    }
    
    \label{fig:fig6}
\end{figure}

We consider our qubit-coupler system operating in the dispersive regime, where qubit-coupler detuning is the following: $g_{jc} \ll \Delta_{jc}$ ($j=1,2$). Qubit-qubit interaction operates in the straddling regime where the qubit detuning $|\Delta_{12}| = |\omega_1-\omega_2| \leq |\eta_{1(2)}|$. $XY$ coupling  (Eq. \ref{eq::Htrans} in the main text) between the qubits is $J = g_{12} + g_{1c}g_{2c}/\Delta$. $ZZ$-coupling occurs from the interaction of qubit states with qubits with non-computational levels: $\ket{101} \leftrightarrow |020\rangle$ and $\ket{101} \leftrightarrow |002\rangle$ ($|200\rangle$). By doing rotating-wave perturbative and deriving up to fourth-order perturbation one can obtain the approximate expression \cite{Zhao_Lan_Xu_Xue_Blank_Tan_Yu_Yu_2021} $\zeta \approx \zeta_{020} + \zeta_{200} + \zeta_{002} + \zeta_1$ with
\begin{align}
    \zeta_{020} &= \frac{J_{020}^2}{\Delta_1 + \Delta_2 - \eta_c}, & \zeta_{200} &= \frac{J_{200}^2}{\Delta_{12} - \eta_2}, \nonumber \\
    \zeta_{002} &= -\frac{J_{002}^2}{\Delta_{12} + \eta_1}, & \zeta_1 &= \frac{4g_{12}g_{1c}g_{2c}}{\Delta_1 \Delta_2},
    \label{eq:A2}
    \tag{A1}
\end{align}
where $\zeta_{020}$, $\zeta_{200}$ and $\zeta_{002}$ are $ZZ$ contributions from the effective coupling of $\ket{101}$ with $\ket{020}$, $\ket{200}$ and $\ket{002}$ levels, respectively, $\zeta_1$ is the contribution from the interaction of the lower levels of the qubit and the coupler. In turn, the effective coupling strengths are defined as:
\begin{align}
    J_{020} &\approx \sqrt{2} g_{1c}g_{2c} \left(\frac{1}{\Delta_1} + \frac{1}{\Delta_2}\right) \approx \frac{2\sqrt{2} g_{1c}g_{2c}}{\Delta}, \nonumber \\
    J_{200} &\approx \sqrt{2} \left(g_{12} + \frac{g_{1c}g_{2c}}{\Delta_1}\right) \approx \sqrt{2} J, \nonumber \\
    J_{002} &\approx \sqrt{2} \left(g_{12} + \frac{g_{1c}g_{2c}}{\Delta_2}\right) \approx \sqrt{2} J,
    \label{eq:A3}
    \tag{A2}
\end{align}

Now consider the main coupler leakage states for the CPhase gate in the single- and two-photon subspaces. During the gate execution, the states $\ket{101}$ and $\ket{002}$ are brought into resonance. $\ket{001}$ becomes detuned from $\ket{100}$ by $\eta_1$. Thus, in the single-photon subspace $\ket{100}$ becomes strongly hybridized with $\ket{010}$ since $\ket{001}$ is detuned on the lower frequency. In the two-photon subspace, similarly, $\ket{101}$ is strongly coupled to $\ket{110}$ and much weaker to $\ket{011}$. The tunable coupler state $\ket{020}$ is coupled to $\ket{101}$ only through a second-order process and the effective interaction is weak, so we neglect $\ket{020}$. We can now write the effective two-state Hamiltonians for the single- and two-photon subspaces:
\begin{align}
    H_1^{\text{CPhase}} = 
    \begin{array}{c}
    \begin{pmatrix}
    \omega_1 & g_{1c} \\
    g_{1c} & \omega_c
    \end{pmatrix} \\[1.5ex]
    \Ket{100} \quad \Ket{010}
    \end{array}
    \equiv
    \begin{array}{c}
    \begin{pmatrix}
    -\Delta_1/2 & g_{1c} \\
    g_{1c} & \Delta_1/2
    \end{pmatrix} \\[1.5ex]
    \Ket{100} \quad \Ket{010}
    \end{array}, 
    \tag{A3}
\end{align}    
\begin{align}
    H_2^{\text{CPhase}} = 
    \begin{array}{c}
    \begin{pmatrix}
    \omega_1+\omega_2 & g_{1c} \\
    g_{1c} & \omega_c+\omega_2
    \end{pmatrix} \\[1.5ex]
    \Ket{101} \quad \Ket{011}
    \end{array}
    \equiv
    \begin{array}{c}
    \begin{pmatrix}
    -\Delta_1/2 & g_{1c} \\
    g_{1c} & \Delta_1/2
    \end{pmatrix} \\[1.5ex]
    \Ket{101} \quad \Ket{011}
    \end{array}.
    \tag{A4}
\end{align}

The angle $\theta$ between the eigenstate vector and the computational state axis is given by $\theta = \arctan(2g_{1c}/\Delta_1)$. Consequently, the overlap between the eigenstate and coupler leakage state becomes:
$O = \sin^2(|\theta|/2).$
Assuming the coupler remains strongly detuned:
$O \approx \frac{|\theta|^2}{4} \approx \frac{g_{1c}^2}{\Delta_1^2},$ 
the effective $|101\rangle$-$|002\rangle$ coupling strength governing diabatic CPhase gate speed approximates as~\cite{Yan_Krantz_Sung_Kjaergaard_Campbell_Orlando_Gustavsson_Oliver_2018}:
 $\tilde{g} \approx \sqrt{2}J \approx \sqrt{2}\left(g_{12} + \frac{g_{1c}g_{2c}}{\Delta_2}\right).$ 
For simplicity, given that $g_{12}$ is much lower than $g_{1c(2c)}$ and $g_{1c} \approx g_{2c}$, the equivalent coupling strength reduces to:
$\tilde{g} \approx \frac{g_{1c}^2}{\Delta}.$

We can now observe that as the coupler detuning decreases during gate operation, the coupling strength increases faster than the eigenstate's overlap with the coupler leakage state. Leveraging this effect, we engineered a system with strong coupler detuning that maintains overlap $O<1\%$ while achieving 25 MHz coupling strength for 20-ns pulses. 

The large coupler detuning provides an additional advantage: complete suppression of residual ZZ coupling at the parking frequency. When detuning is sufficiently large, the $\zeta_{020}$ and $\zeta_1$ terms contributing to ZZ coupling become negligible. In this regime, the ZZ interaction is dominated solely by higher qubit level couplings $\zeta_{200(002)} \propto (\sqrt{2}J)^2$, meaning it is simultaneously suppressed with the XY coupling. 

Numerical simulations confirm the concurrent suppression of both XY and ZZ interactions at parking frequencies, while maintaining minimal state overlap $O<1\%$ with strong effective coupling in the strongly-detuned coupler configuration. Fig. \ref{fig:fig5}\textbf{a},\textbf{b} show numerically simulated maps of effective XY coupling strength $\tilde{g}/2\pi$ and residual ZZ interaction $\zeta/2\pi$ as functions of both $g_{12}$ and tunable coupler frequency. The qubits are resonantly tuned to 5 GHz which is a common situation at the idle operation. All circuit parameters are the same as for the experimental device (Table~1 in main text). 

For $g_{12}=15$ MHz as in our experimental device the coupler requires high frequency ($> 10$ GHz) to turn off the coupling, with residual ZZ interaction dominated by $\zeta_{200}$ and $\zeta_{002}$ terms. The ZZ switch-off frequencies degenerate into one frequency, which is the same as for the XY coupling. 

Fig. \ref{fig:fig5}\textbf{c},\textbf{d} show the numerically simulated overlaps of computational states with leakage states and the coupling strength $\tilde{g}_{\text{CZ}}$ as a function of the coupler frequency. We used our experimental device parameters with Qubit~1 fixed at 5 GHz and Qubit~2 detuned by the anharmonicity, matching the actual CZ gate conditions. We calculated the overlap between dressed states $\widetilde{|100\rangle}$ and $\widetilde{|101\rangle}$ with leakage states $\widetilde{|010\rangle}$ and $\widetilde{|110\rangle}$ ($\widetilde{|011\rangle}$) for the varying coupler frequencies. The states $\widetilde{|100\rangle}$ and $\widetilde{|101\rangle}$ represent the system's eigenstates when the coupler is tuned to the frequency that turns off the effective coupling: $\tilde{g}_{\text{CZ}}=0$. 

The leakage state overlap grows slower than the coupling strength as coupler detuning decreases. At coupling strengths sufficient for CZ gate consisting of two 20-ns pulses, the state overlap coefficients remain below 1\%.

Fig. \ref{fig:fig6}\textbf{a},\textbf{b} show the resulting engineered energy level diagram for the 2-qubit device, which incorporates all the features described above.

\section{CPhase calibration procedure}
\label{app:C}

The calibration procedure of BAT pulse sequence is straightforward (Fig. \ref{fig:fig7_calib}\textbf{a,b}). In the first step, the goal is to calibrate the pulse amplitudes on the qubit and the coupler that enable a swap between the $\ket{101}$ and $\ket{002}(\ket{200})$ states. We excite both qubits to prepare the $\ket{101}$ state. Then, we apply pulses to the qubit and the coupler. At the end of the sequence both qubits are readout.
First, we sweep the coupler amplitude (Fig. \ref{fig:fig7_calib}\textbf{c,e})., and then the qubit amplitude (Fig. \ref{fig:fig7_calib}\textbf{d,f})., in order to minimize the $\ket{101}$ population and maximize the $\ket{002}(\ket{200})$ population, respectively 
Next, a sequence consisting of an odd number N of pulses is applied, with a 20 ns delay after each pulse. In our experiments, we performed sequences up to N = 101.
The sequences are executed iteratively. At each iteration, the the amplitude is updated by selecting the optimal value.
The procedure is first performed for one pulse polarity and then repeated for the other.

\begin{figure}
    \centering
    \includegraphics[width=1\linewidth]{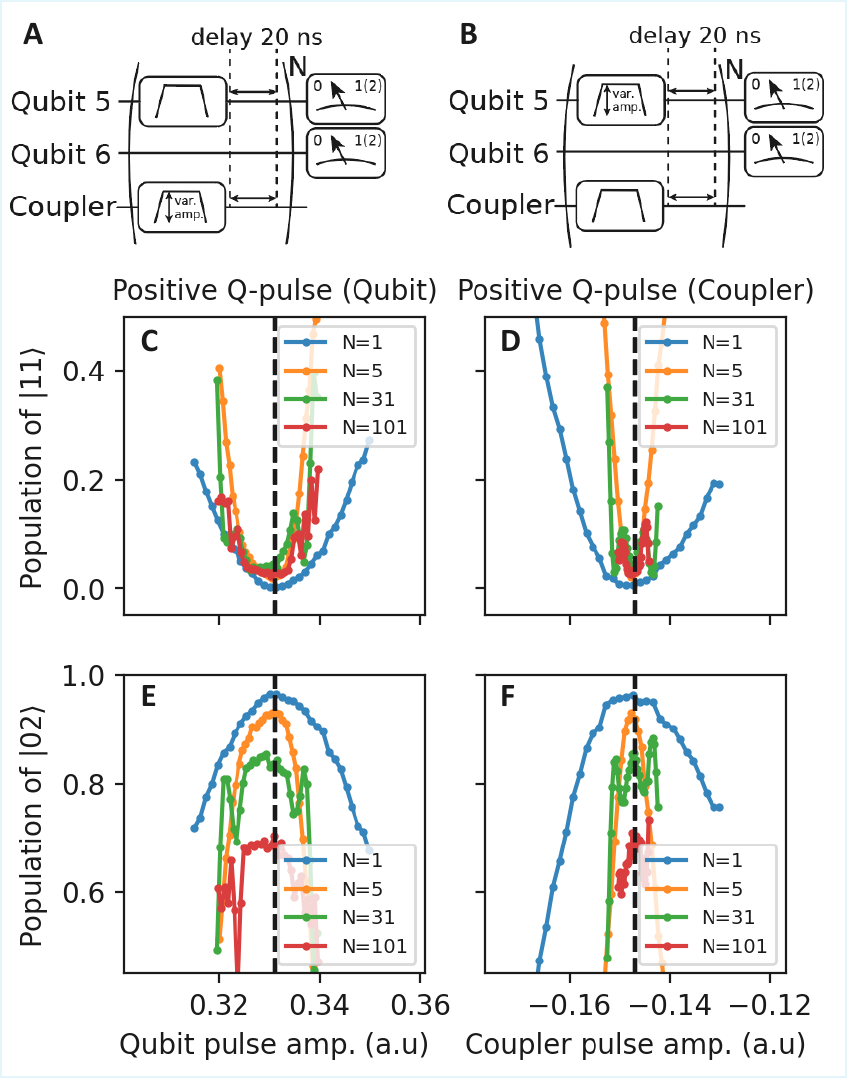}
    \caption{ Calibrations for the CPhase gate of the Q5-Q6 qubit pair of the 8-qubit device.
    \textbf{(a)} Pulse sequences for the calibration of the coupler pulse amplitude and
    \textbf{(b)} qubit pulse amplitude for the CPhase gate. Measured population of $\ket{11}$ versus \textbf{(c)} qubit pulse amplitude and \textbf{(d)} coupler pulse amplitude for different number of pulses N. Measured population of $\ket{02}$ versus \textbf{(e)} qubit pulse amplitude and \textbf{(f)} coupler pulse amplitude.
    }
    
    \label{fig:fig7_calib}
\end{figure}

The goal of the second calibration step is to set the desired conditional phase $\phi_{2Q}$. A pulse sequence consisting of two pulses of opposite polarities with a small pulse in between is constructed.(Fig. \ref{fig:fig8_calib}\textbf{a}) The amplitude of the small pulse $a_\text{int}$ and the delay after it are swept (Fig. \ref{fig:fig8_calib}\textbf{b}). Quantum state tomography is performed to determine the conditional phase.
By jointly varying the amplitude of the small pulse and the delay, arbitrary conditional phase $\phi_{2Q}$ values can be achieved while keeping the amplitude of $a_\text{int}$ sufficiently small to avoid residual tailing effects after the pulses.
Next, the amplitude of the small pulse is fine-tuned to obtain the target phase (Fig. \ref{fig:fig8_calib}\textbf{c}), in this case, $\phi_{2Q}=180$ deg. To do this, the amplitude is swept for sequences consisting of an odd number N of gates. In our experiments, we performed sequences up to N = 5. The intersection point of the resulting curves corresponds to the precisely calibrated CZ amplitude.
This calibration approach can similarly be applied to other target conditional phases by choosing a different N.

\begin{figure}
    \centering
    \includegraphics[width=1\linewidth]{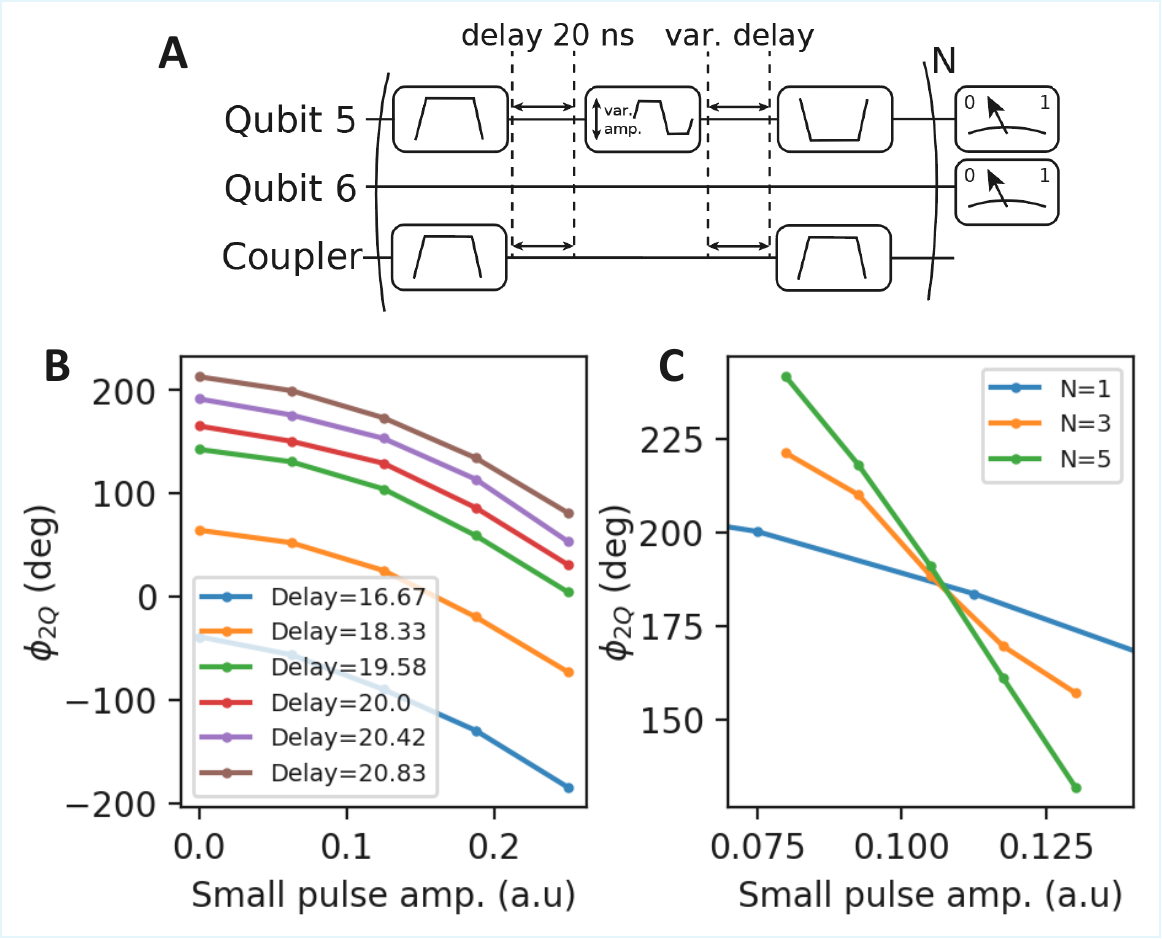}
    \caption{ Calibrations of the conditional phase of the Q5-Q6 qubit pair of the 8-qubit device.
    \textbf{(a)} Pulse sequences for the calibration of the small pulse amplitude $a_\text{int}$.
    \textbf{(b)} Measured conditional phase $\phi_{2Q}$ versus delay and small pulse amplitude $a_\text{int}$. \textbf{(c)} Fine small pulse amplitude tuning for achieving $\phi_{2Q}=180$ deg. Measured conditional phase $\phi_{2Q}$ versus small pulse amplitude $a_\text{int}$ for N gates in the sequence. The intersection point of the curves corresponds to precisely calibrated small pulse amplitude for CZ gate. 
    }
    
    \label{fig:fig8_calib}
\end{figure}

\section{8-qubit quantum processor}
\label{app:D}

\begin{table*}

    \centering
    
    \caption{Performance parameters for the 8-qubit superconducting processor}
    \label{tab:8qperformance}
    \begin{tabular*}
    {\textwidth}{@{\extracolsep{\fill}}lcccccccc@{}}
\toprule
        Parameter & Q1 & Q2 & Q3 & Q4 & Q5 & Q6 & Q7 & Q8 \\[2pt]
\midrule
        $\omega_i/2\pi$ (GHz) & 5.288 & 5.361 & 5.207 & 5.261 & 5.156 & 5.239 & 5.055 & 5.076 \\[2pt]
        
        $\eta_i/2\pi$ (GHz) & -0.241 & -0.246 & -0.239 & -0.245 & -0.242 & -0.247 & -0.239 & -0.243 \\[5pt]

       $\zeta/2\pi$ (kHz) & \shortstack{3.2\\ (Q1-Q3)} & \shortstack{3.2\\ (Q2-Q3)} & \shortstack{2.3\\ (Q3-Q6)} & \shortstack{1.1\\ (Q4-Q3)} & \shortstack{9.6\\ (Q5-Q6)} &  & \shortstack{2.1\\ (Q5-Q6)} & \shortstack{1.0\\ (Q8-Q6)} \\[5pt]

       $T_{1,idle}$ (\si{\micro s}) & 29.2 & 34.5 & 20.27 & 18.45 & 18.50 & 38.3 & 27.8 & 48.2 \\[3pt]

       $T_{2E,idle}$ (\si{\micro s}) & 41.9 & 20.7 & 31.2 & 30.2 & 32.4 & 28.5 & 43.1 & 51.1 \\[3pt]

       $T_{2R,idle}$ (\si{\micro s}) & 16.8 & 4.4 & 15.3 & 13.7 & 10.1 & 5.2 & 5.7 & 50.3 \\[3pt]
        
        $F_{1Q,iso}$ $(\%)$ & 99.944(2) & 99.937(4) & 99.931(2) & 99.934(4) & 99.950(4) & 99.928(7) & 99.953(4) & 99.962(2) \\[3pt]

        $F_{1Q,simult}$ $(\%)$ & 99.897(7) & 99.919(6) & 99.926(4) & 99.919(4) & 99.878(9) & 99.924(3) & 99.952(3) & 99.953(3) \\[5pt]

        $F_{CZ}$ $(\%)$ & \shortstack{98.72(0.2)\\ (Q1-Q3)} & \shortstack{99.33(9)\\ (Q2-Q3)} & \shortstack{99.08(0.12)\\ (Q3-Q6)} & \shortstack{99.09(0.16)\\ (Q3-Q4)} & \shortstack{99.47(9)\\ (Q5-Q6)} &  & \shortstack{99.37(0.13)\\ (Q7-Q6)} & \shortstack{99.18(0.19)\\ (Q8-Q6)} \\[5pt]
        
        $F_{R0}$ $(\%)$ & 99.62 & 99.64 & 99.23 & 99.02 & 99.36 & 98.98 & 99.33 & 99.35 \\
        
        $F_{R1}$ $(\%)$ & 97.52 & 96.91 & 96.38 & 95.90 & 96.35 & 95.40 & 96.02 & 95.03 \\

    \bottomrule
    \end{tabular*}
    
    \label{tab:placeholder}
\end{table*}    

We demonstrate the potential for further scaling using the proposed architecture and our method for implementing two-qubit gates on an 8-qubit quantum processor. Processor architecture is similar to our 4-qubit device employing BAT pulses. Each qubit is connected to an individual microwave control line. The 25-ns single-qubit gates are calibrated using the DRAG method. Qubits and couplers are connected to dedicated flux control lines. CZ gates were calibrated in the same manner as in the 4-qubit and 2-qubit device. Readout is performed via quarter-wave resonators coupled to a common bandpass Purcell filter. The processor connectivity scheme is shown in Fig. \ref{fig:fig9_8q}\textbf{a}. The parameters of the 8-qubit processor are summarized in Table \ref{tab:8qperformance}.

\begin{figure*}
    \centering
    \includegraphics[width=0.92\linewidth]{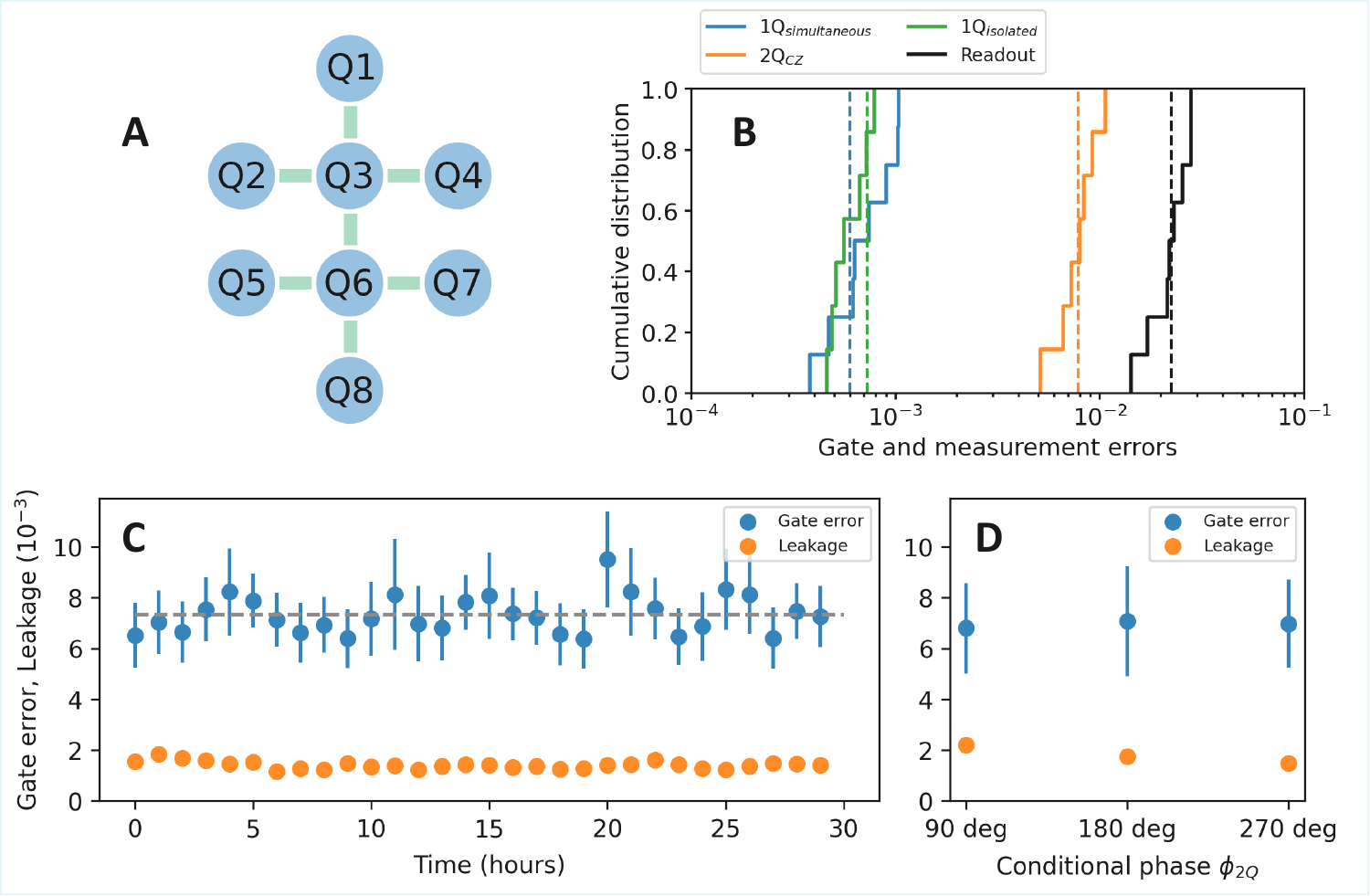}
    \caption{ 
    \textbf{(a)} Connectivity scheme for the 8-qubit quantum processor.
    \textbf{(b)} Cumulative distributions of the operation errors of the 8-qubit processor: simultaneous single-qubit gates (blue), isolated single-qubit gates (green),two-qubit CZ gates (orange), two-state readout (black). Dashed vertical lines represent average values. \textbf{(c)} CZ gate error and leakage as a function of time of the Q5-Q6 qubit pair. Mean gate error is shown as gray dashed line. \textbf{(d)} CZ error and leakage for the Q5-Q6 qubit pair for various calibrated conditional phases $\phi_{2Q}$. The uncertainty represents the 68\% confidence interval. The error bars for leakage are smaller than the data points.
    }
    
    \label{fig:fig9_8q}
\end{figure*}

The average fidelity of single-qubit gates performed simultaneously is $99.93(2)\%$, compared to $99.94(1)\%$ for isolated gates. The average two-qubit CZ gate fidelity is $99.21(0.16)\%$, with a maximum of $99.49\%$. The average two-state readout fidelity is $97.75(0.45)\%$. Cumulative distribution of the processor operation errors is shown in Fig. \ref{fig:fig9_8q}\textbf{b}. The residual ZZ coupling $\zeta/2\pi$ is below 10 kHz for all coupled pair of qubits. The reported readout fidelity corresponds to two-state readout.

We used a Q5-Q6 pair on an 8-qubit processor to test the stability of the CZ gate calibration using the proposed scheme (Fig. \ref{fig:fig9_8q}\textbf{c}). We calibrated the gate as described in Appendix C and ran long-term XEB measurements, triggered every hour, obtaining data on calibration stability over 30 hours. We observed stable gate fidelity with a mean of $99.27(7)\%$ and leakage of $0.14(1)\%$ with no significant drift.

We then used the same qubit pair to evaluate the gate fidelity for various conditional phases (Fig. \ref{fig:fig9_8q}\textbf{d}). We separately calibrated gates for $\phi_{2Q}=90$ deg., 180 deg. and 270 deg. and measured their fidelities using XEB. We obtained nearly identical fidelities and leakage across the entire set of angles.

In addition, we compared the performance of the bipolar and unipolar pulse schemes on the Q5-Q6 pair. We calibrated the CZ gate and performed XEB testing for both cases, obtaining two-qubit CZ gate fidelities of $99.29(9)\%$ for bipolar pulses and $99.01(14)\%$ for unipolar pulses (Fig. \ref{fig:fig10_univsbi}\textbf{a}). The extracted average reference fidelity is $99.81(2)\%$ The average leakage was nearly identical: $0.16(1)\%$ for unipolar pulses and $0.16(1)\%$ for bipolar pulses (Fig. \ref{fig:fig10_univsbi}\textbf{b}). The average purity error per cycle was $0.61(16)\%$ for the bipolar gates versus $0.89(23)\%$ for the unipolar, suggesting that the main difference in fidelity between the two schemes is due to decoherence errors. We attribute this to the fact that bipolar pulses provide additional echo-like protection against low-frequency noise. Average purity error of the reference cycle consisting only of single-qubit operations is $0.13(2)\%$ for comparison.

\begin{figure}
    \centering
    \includegraphics[width=0.9\linewidth]{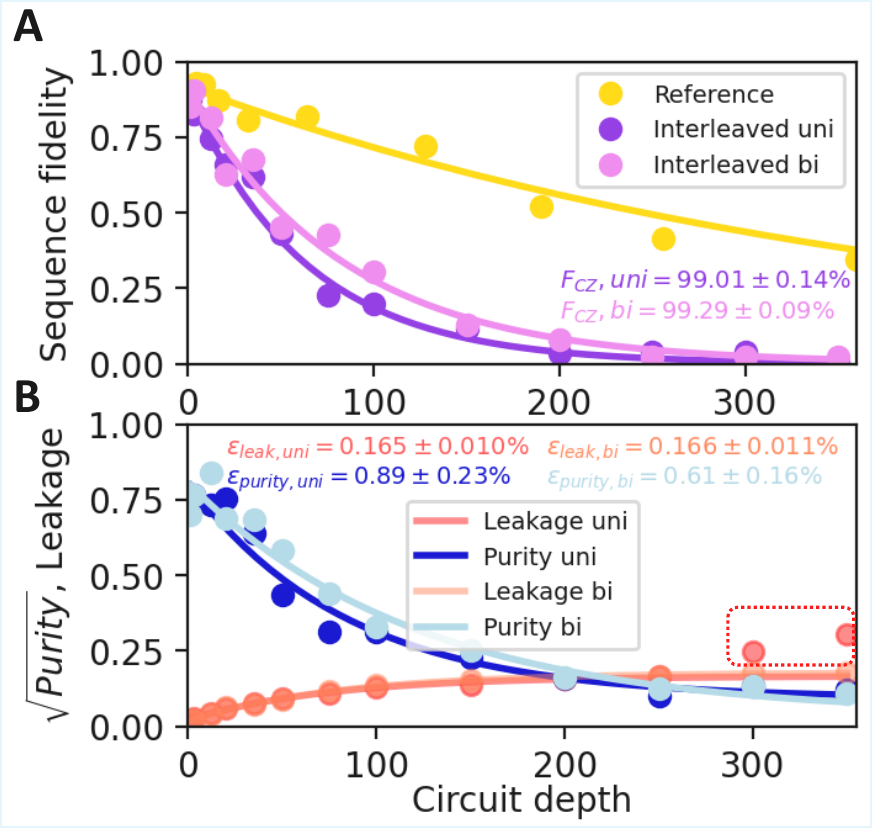}
    \caption{ Comparison of unipolar and bipolar pulses performed on the Q5-Q6 qubit pair:
    \textbf{(a)} The results of interleaved XEB for the 100-ns-long CPhase gate, consisting of two 20-ns pulses unipolar pulses (dark violet) versus bipolar pulses (violet). 
    \textbf{(b)} Accumulation of population in the leakage subspace (red and pink for bipolar and unipolar pulses respectively) and decay of square root of purity during interleaved XEB (light blue and dark blue for bipolar and unipolar pulses respectively). Leakage for unipolar pulses was fitted using the data excluding circuit depths of 300 and 350. For each circuit depth, we sample over 30 different random circuits.The probability distributions were obtained from 4096 single-shot readouts. The uncertainty represents the 68\% confidence interval.  }
    
    \label{fig:fig10_univsbi}
\end{figure}
The main drawback of the unipolar pulses in our scheme manifests itself as an arbitrary waveform generator overrange due to qubit pulse precompensation. After a circuit depth of 300 in XEB, the generator exceeded its voltage range, leading to deviations in the qubit pulse amplitudes from their calibrated values and a sharp increase in leakage. In Fig. \ref{fig:fig10_univsbi}\textbf{b} the points with sharply increased leakage deviating from the fit line are highlighted with a red dashed frame.

\section{Single-qubit gate fidelities}

\label{app:E}

The measurement of single-qubit gate fidelity was performed using cross-entropy benchmarking. Each cycle consists of X and Y rotations with angles randomly sampled  between $-\pi$ and $\pi$. We generate 30 random sequences of up to 2000 cycles and measure in the Z basis using single-shot readout with 4096 acquisitions.

\begin{figure}
    \centering
    \includegraphics[width=1\linewidth]{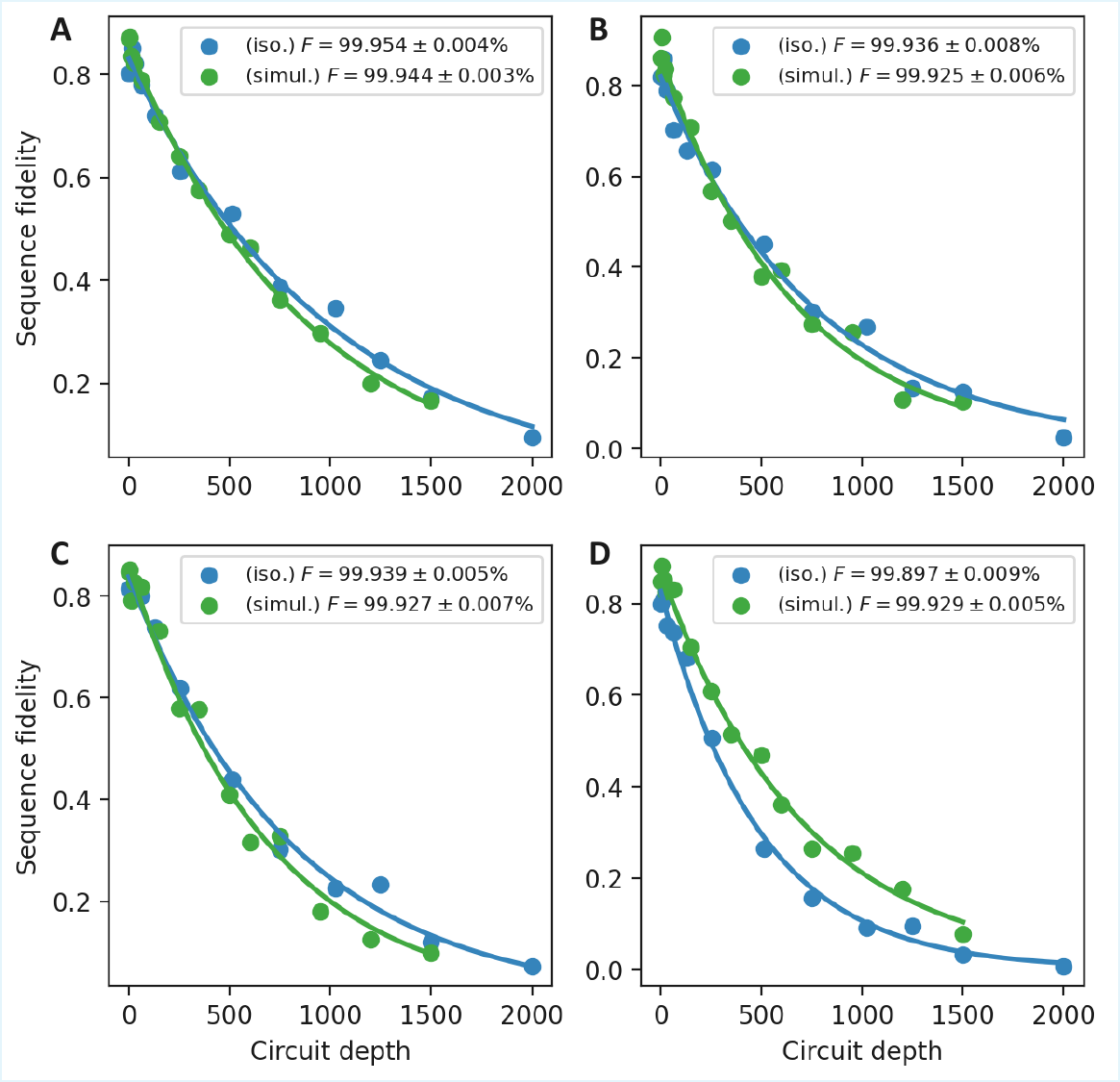}
    \caption{ Cross-entropy benchmarking (XEB) of single-qubit gates of:  
    \textbf{(a)} qubit 1,  
    \textbf{(b)} qubit 2,
    \textbf{(c)} qubit 3 and
    \textbf{(d)} qubit 4. The data points show the sequence fidelity, and the line is an exponential fit to the data. The legend indicates the fidelities when qubits are driven isolated (blue) and simultaneous (green).}
    \label{fig:fig_SQ}
\end{figure}

The results of the single-qubit XEB of the 4-qubit processor are shown in Fig. \ref{fig:fig_SQ}. We performed benchmarking both on isolated qubits and under simultaneous qubit drive. The qubits remained at their idle frequencies (see Table \ref{tab:processor} in the main text) in both cases.
We observe a slight decrease in fidelity during simultaneous benchmarking on qubits 1, 2, and 3. We attribute this to microwave crosstalk or residual  coupling. Meanwhile, we suspect that the fidelity for qubit 4 was lower in the isolated test than in the simultaneous test due to possible fluctuations in coherence during testing, for instance, caused by a two-level defect.

\section{Device fabrication and measurement setup}
\label{app:F}

\begin{figure}
    \centering
    \includegraphics[width=1\linewidth]{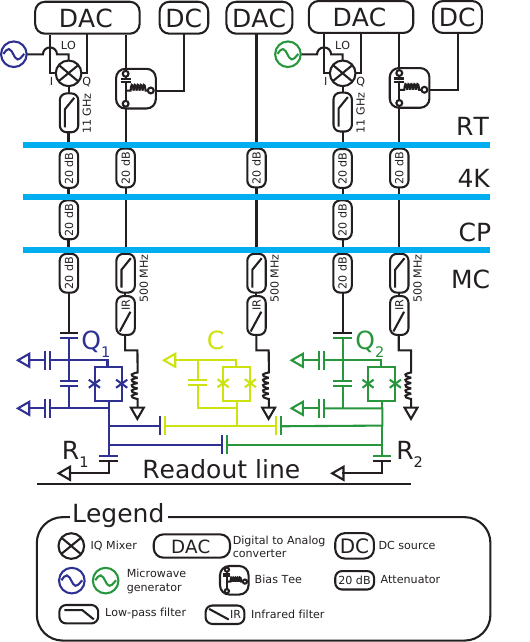}
    \caption{A schematic diagram of the experimental setup for the 2-qubit unit cell.
    }
    
    \label{fig:fig7}
\end{figure}

The samples used in the experiments are fabricated using high-resistivity silicon substrates (\SI{10000}{\ohm\centi\meter}). Prior to the base layer deposition, the substrate is cleaned in a Piranha solution at \SI{80}{\celsius}, followed by dipping in hydrofluoric bath. \SI{100}{\nano\meter}-thick Al base layer is deposited using ultra high vacuum e-beam evaporation system. Pads were defined using a direct-laser lithography and wet-etched. Floating qubit design is similar to Ref.~\cite{Smirnov_Krivko_Solovyova_Ivanov_Rodionov_2024}.

The Josephson junctions were fabricated using Dolan technique as described in Refs.~\cite{moskalev2023optimization,Pishchimova_Smirnov_Ezenkova_Krivko_Zikiy_Moskalev_Ivanov_Korshakov_Rodionov_2023}. The substrate is spin coated with resist bilayer composed of \SI{500}{\nano\meter} EL9 copolymer and \SI{100}{\nano\meter} CSAR 62. Layouts were generated and exposed with \SI{50}{\kilo\electronvolt} e-beam lithography system. The development was performed in a bath of amylacetate followed by IPA dip and additional in IPA/DIW solution. Al/AlOx/Al junctions are shadow evaporated in ultra-high vacuum deposition system. Resist lift-off was performed in N-methyl-2-pyrrolidone at \SI{80}{\celsius}.

Then aluminum bandages are defined and evaporated using the same process as for the junctions with an in-situ Ar ion-milling to provide good electrical contact of the junction with the base layer. Lift-off is performed in a bath of N-methyl-2-pyrrolidone with sonication at \SI{80}{\celsius} and rinsed in a bath of IPA with ultrasonication. Finally, \SI{300}{\nano\meter}-thick aluminum air bridges are fabricated by sputtering and subsequent wet etch as in Ref.~\cite{zikiy2023high}

The detailed experimental qubit control setup scheme is shown in Fig.~\ref{fig:fig7}. The devices are measured in a Bluefors LD400 dilution refrigerator. One line connected to the chip is used for readout and the others for applying single qubit gates (XY controls) and frequency control (Z). Pulsed XY control of the qubits was realized by upconverting the intermediate-frequency in-phase and quadrature signals from the digital analog converter (DAC) of the arbitrary waveform generator (AWG), using IQ-mixer and microwave local oscillator. 

To control the frequency of each qubit, a current source at room temperature is connected to a DAC output using the bias tees. Couplers are controlled using DAC-generated pulses only. The \SI{20}{dB} attenuators are thermalized at the corresponding temperature stages, as shown in Fig.~\ref{fig:fig7}. 

Readout tone was generated by DAC and up-converted to the readout resonator frequency using mixer and microwave local oscillator. Readout microwave signal passed through the chip, is amplified by a cryogenic impedance matched parametric amplifier~\cite{moskaleva2024lumped} (IMPA) and then downconverted. The readout signal is also amplified by high-electron mobility transistors at the \SI{4}{K} stage of the cryostat and at room temperature. 

We use a DC source to tune-up the parametric amplifier to the desired frequency and pump the IMPA by microwave source of vector network analyzer. Readout lines and flux control lines are additionally equipped with custom-made infrared filter on the cryostat mixing stage to suppress infrared noise and standing waves. Sample holders with IMPA and qubit chips are placed in the magnetic shields.


\end{acknowledgments}




\bibliography{Couplers.bib}

\end{document}